\documentclass[aps,prb,footinbib,onecolumn,notitlepage]{revtex4-2}
\usepackage[a4paper,top=2cm,bottom=2cm,left=1.75cm,right=1.75cm]{geometry}
\usepackage{amsmath}
\usepackage{amssymb}
\usepackage{color}
\usepackage{graphicx}
\usepackage{ifpdf}
\usepackage{epstopdf}
\DeclareGraphicsExtensions{.eps, .pdf, .jpg, .tif}
\usepackage{hyperref}
\usepackage{bm}
\usepackage{enumerate}
\usepackage{nicefrac}
\setcitestyle{numbers} 
\def\ns{\negthickspace}
\def\vj{{\bf j}} 
 
\def\vp{{\bf p}} 
\def\vr{{\bf r}} 
 
\def\vA{{\bf A}} 
\def\vB{{\bf B}}

\newcommand{\grad}{\pmb\nabla}
\newcommand{\be}{\begin{equation}}
\newcommand{\ee}{\end{equation}}
\newcommand{\ber}{\begin{eqnarray}}
\newcommand{\eer}{\end{eqnarray}}
\begin{document}
\title{Photon Frequency Conversion in High-$Q$ Superconducting Resonators: \\
\emph{Axion Electrodynamics, QED \& Nonlinear Meissner Radiation}}
\author{Hikaru Ueki and J. A. Sauls}
\affiliation{Hearne Institute of Theoretical Physics,
	     Louisiana State University, Baton Rouge, LA 70803, USA}
\date{\today}

\begin{abstract}
High-Q superconducting resonators have been proposed and developed as detectors of light-by-light scattering mediated by the hypothesized axion or virtual electron-positron pairs in quantum electrodynamics - the Euler-Heisenberg (EH) interaction.
Photon frequency and mode conversion is central to the scheme for detecting such rare events.
Superconducting resonators are nonlinear devices. The Meissner screening currents that confine the electromagnetic fields to the vacuum region of a superconducting RF cavity are nonlinear functions of the EM field at the vacuum-superconducting interface, and as a result can generate source currents and frequency conversion of microwave photons in the cavity.
In this report we consider photon frequency and mode conversion in superconducting resonators with high quality factors from Meissner currents in single and dual cavity setups proposed for axion and QED searches based on light-by-light scattering.
In a single cavity with two pump modes photon frequency conversion by the Meissner screening current dominates photon generation by the EH interaction for cavities with $Q \lesssim 10^{12}$. The Meissner currents also generate background photons that limits the operation of the resonator for axion detection in three-mode, single cavity setups. We also consider the leakage of photons from pump modes into the signal mode for both axion and EH mediated light-by-light scattering.
Photon frequency conversion by the EH interaction can compete with Meissner and leakage radiation in \emph{ultra-high-Q} cavities that are beyond current state of the art.
Meissner radiation and leakage backgrounds can be suppressed in dual cavity setups with appropriate choices for pump and spectator modes, as well as the single-cavity setup proposed for heterodyne detection of galactic axion dark matter.
\end{abstract}

\maketitle

\section{Introduction} 

The axion was proposed as an elegant solution to the strong CP problem~\cite{pec77,wil78,wei78}. It was later realized that if it exists it would be a viable candidate for dark matter over a wide range of possible axion masses~\cite{pre83,abb83,din83}.
The pseudo-scalar axion field has a symmetry allowed coupling to the electromagnetic field described by the interaction Lagrangian~\footnote{We follow as much as possible the notation of Ref.~\cite{kah22}.}, 
${\cal L}_{a}^{\rm int} 
= -\frac{1}{4}\,g_{a\gamma\gamma}\,a\,F_{\mu\nu}\tilde{F}^{\mu\nu}
= g_{a\gamma\gamma}\,a\,{\bf E}\cdot{\bf B}$,
where $a$ is axion field,
$F_{\mu\nu}$
is the EM field tensor, $\tilde{F}^{\mu\nu}=\epsilon^{\mu\nu\lambda\rho}F_{\lambda\rho}$ is the dual EM field tensor, and 
${\bf E}\cdot{\bf B}$
is the pseudo-scalar constructed from the EM fields. 
The axion-photon coupling is $g_{a\gamma\gamma}$. This interaction, and the corresponding axion sources to Maxwell's equations, spawned proposals for detecting axions or the axion field~\cite{sik83}. For a recent review of ongoing and proposed experimental searches c.f. Ref.~\cite{ada22}.

Here we consider recent proposals for axion detection based on Superconducting Radio-Frequency (SRF) cavities in the absence of static magnetic fields. Niobium SRF cavities have been developed with the quality factors in the range $Q\sim10^9-10^{11}$ for the resonant frequencies $f\sim50 \ {\rm MHz}-4$ GHz, and can support field energy densities as high as $N\approx 10^{25}$ photons per mode~\cite{pad17,gra13,rom14}. These cavities, developed for high-energy particle accelerators~\cite{padamsee08,gur17}, are now being developed as detectors to capture physics beyond the Standard Model, nonlinear QED corrections to the free electromagnetic (EM) field, as well as gravitational wave (GW) detectors~\cite{bog19,ber20,gao21,ber21,kah22,rom23,ber23}.
High-Q cavities are also nonlinear superconducting devices that can be used to study novel nonequilibrium and nonlinear superconducting phenomena, including microwave photon generation~\cite{sau22}, which is central to the analysis and results reported here.

\subsection{Photon-Photon Scattering via the Axion Field} 

\begin{figure*}[t]
\begin{center}
\includegraphics[width=0.8\linewidth]{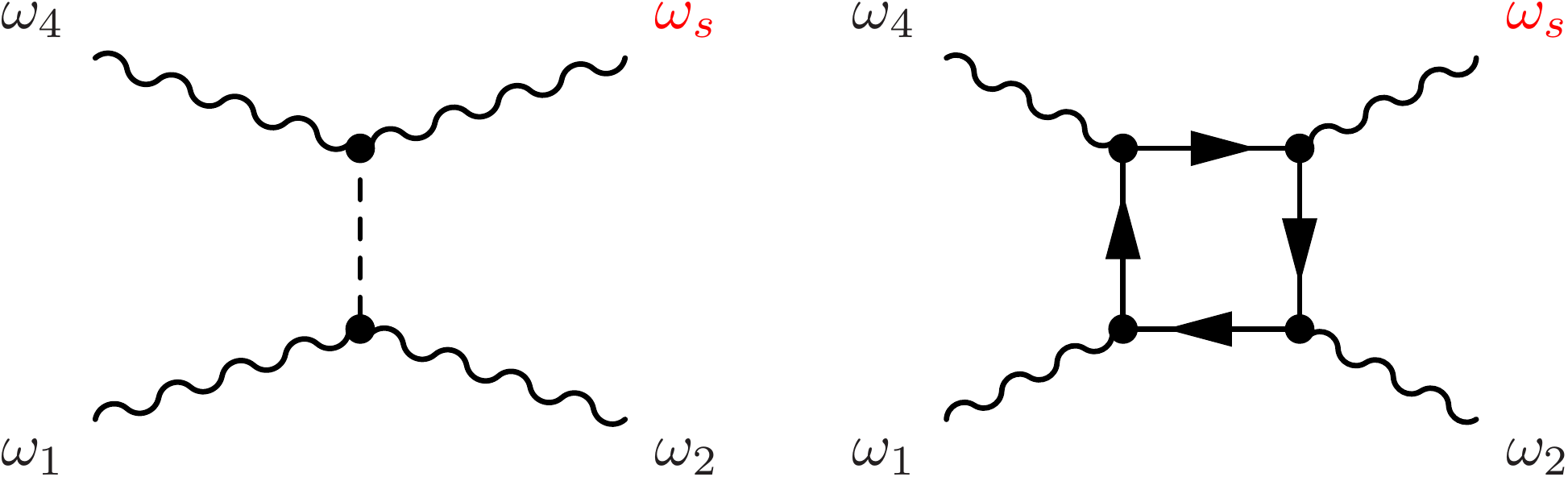}
\caption{Feynman diagrams for the light-by-light scattering mediated by the axion field (left diagram) and electron-positron pairs (right diagram), where the wavy, dashed, and solid lines denote photon, axion, and electron, respectively. 
Two modes are pumped with photons with frequencies $\omega_1$ and $\omega_2$, while signal photons are generated a the intermodulation frequency $\omega_s=2\omega_1-\omega_2$, designed to be a resonant mode of the cavity.}
\label{fig-feynman_diagrams}
\end{center}
\end{figure*}

To leading order in the axion-EM coupling, virtual axions mediate a quartic interaction between photons governed by the effective Lagrangian~\cite{sik83,bog19}
\begin{equation} 
{\cal L}^{{\rm eff}}_{a}
=
\frac{g_{a\gamma\gamma}^2}{128\pi}\frac{\hbar^2}{m_a^2c^2}
(F_{\mu\nu}\tilde{F}^{\mu\nu})^2
\,,
\end{equation}
for photon energies, $\hbar\omega_{\gamma}\ll m_a c^2$, where $m_a$ is the axion mass. This interaction describes photon-photon scattering mediated by the axion field as shown in Fig.~\ref{fig-feynman_diagrams}~\cite{kah22}.

Bogorad \emph{et al.} proposed using three resonant modes of a single high-Q SRF cavity as a platform for detecting the axion field~\cite{bog19}. An alternative proposal by Gao \emph{et al.} is based on an emitter and receiver cavity that requires only two resonant modes, but identical mode spectra for both cavities~\cite{gao21}. A single cavity setup was also proposed by Berlin \emph{et al.} to detect galactic axion dark matter (DM) by observing axion induced transitions of microwave photons between two nearly degenerate resonant modes.
These setups take advantage of the high quality factor of SRF cavities to enhance the number of signal photons. The single cavity setup eliminates the need to engineer \emph{identical} emmitter and receiver cavities.

For the single cavity setup of Ref.~\cite{bog19} the idea is to use the cubic nonlinearity of axion-electrodynamics to detect the axion field by measuring photons at a signal frequency $\omega_s=2\omega_1-\omega_2$ in an SRF cavity simultaneously pumped with photons at two resonant frequencies $\omega_1$ and $\omega_2$. The pump modes are chosen to optimize ${\bf E}\cdot{\bf B}\ne 0$ in the volume of the cavity. Amplification of the signal is ensured by designing the cavity such that the signal frequency is a resonant mode of the cavity.

\subsection{Photon-Photon Scattering \`a la QED} 

Bogorad \emph{et al.} also proposed this platform to detect photon-photon scattering mediated by virtual electron-positron pairs as the perturbative QED correction to the free-field Maxwell equations predicted early on by Euler and Heisenberg (EH)~\cite{hei36} (c.f. Fig.~\ref{fig-feynman_diagrams}). The corresponding effective Lagrangian is~\cite{sch51,bog19,kah22},
\begin{equation}
\hspace*{-2mm}
{\cal L}_{\rm EH}
\ns=\ns
\frac{1}{128\pi}\frac{\hbar e^4}{45\pi m_e^4c^7}
\left[
4(F_{\mu\nu}F^{\mu\nu})^2
\ns+\ns
7(F_{\mu\nu}\tilde{F}^{\mu\nu})^2
\right].
\end{equation}
The Euler-Heisenberg Lagrangian also generates cubic nonlinear corrections to the free-field Maxwell equations via both quartic invariants of the EM field,  
$(F_{\mu\nu}F^{\mu\nu})^2 \propto (|{\bf E}|^2 - |{\bf B}|^2)^2$ 
as well as
$(F_{\mu\nu}\tilde{F}^{\mu\nu})^2 \propto ({\bf E}\cdot{\bf B})^2$.

Thus, for pump modes chosen for axion detection, ${\bf E}\cdot{\bf B}\ne 0$, there are also photons expected to be generated at the signal frequency from the EH interaction. Alternatively, one can choose pump modes with ${\bf E}\cdot{\bf B}=0$ in order to directly measure photon-photon scattering via the EH interaction. Such a measurement would not only confirm the EH prediction of light-by-light scattering, but also provide a proof of principle for discovery of rare events using high-Q SRF cavities as detectors.

\subsection{Photon Frequency Conversion by Meissner Screening Currents} 

With any proposal for detection of rare events the possibility of false positives has to be considered, such as the thermal background of photons at the signal frequency, mechanical vibrations, or leakeage of photons from the pump mode resonances into the signal resonance~\cite{bog19,ber20,gao21,ber21}.
Here we discuss the generation of photons at the signal frequency that arise from the screening currents that confine the EM fields within the SRF cavity.

In contrast to ``perfect conductors'' for which the EM field is excluded from the conductor, the EM field penetrates into a superconductor a distance of order the London penetration depth, $\lambda_{\rm L}\approx 50\,\mbox{nm}$ in Nb. The surface field generates a supercurrent that screens the field from the bulk of the superconductor. 
The Meissner screening current is a function of the gauge-invariant condensate momentum per electron, 
\be
\vp_s=\frac{\hbar}{2}\left(\grad\vartheta - \frac{2e}{\hbar c}\vA\right)
\,, 
\ee
where $\vartheta$ is the phase of the condensate amplitude and $\vA$ is the vector potential. 
For weak surface EM fields, $B\ll B_s\approx 200\,mT$ the Meissner current is approximately linear in EM field, $\vj_s\approx \frac{e}{m^*}\,n_s\,\vp_s$, where $n_s$ is the zero-field superfluid fraction, $e$ is the electron charge and $m^*$ the mass of the conduction electrons in the normal metal. 
This leads to the well known London equation for $\vB=\grad\times\vA$ and screening of the field from the bulk of the superconductor on the scale of the London penetration depth defined by, $\lambda_{\rm L}^{-2}=4\pi n_s\,e^2/m^* c^2$.
However, the Meissner current is in general a nonlinear function of the condensate momentum, and thus the EM field at the vacuum-superconductor interface, 
\be\label{eq-Current_Nonlinear_Perturbative}
\vj_s=
\frac{e}{m^*}\,n_s\,
\left[1-\theta\left(\frac{p_s}{p_{c}}\right)^2\right]\,\vp_s 
\,.
\ee
The magnitude of the nonlinear correction to the screening current is determined by the dimensionless parameter, $\theta$, and the critical momentum set by the gap and Fermi velocity, $p_c=\Delta/v_f$. These parameters depend on temperature and disorder, the latter parametrized by the electron-impurity scattering rate, $1/\tau$, c.f. Ref.~\cite{sau22}.
The nonlinear Meissner (NLM) current has been studied in many different superconducting materials, both theoretically~\cite{yip92a,xu95,dah97,sau22} and experimentally~\cite{oat04,gro10,mak22,wil22}.
In the context of SRF cavities the nonlinear field dependence of the Meissner current becomes important for cavity modes with large field energy density.

Eq.~\eqref{eq-Current_Nonlinear_Perturbative} can be extended to finite frequency at low temperature provided microwave photons do not break pairs, i.e. $\hbar\omega\ll 2\Delta$, which for Nb is $2\Delta/h\simeq 700\,\mbox{GHz}$. Thus at GHz frequencies the nonlinear screening current at a vacuum-superconducting interface is to good approximation given by Eq.~\eqref{eq-Current_Nonlinear_Perturbative} with $n_s$ and $\theta$ given by their static limits. Furthermore, the dissipative component of the current response from thermally excited quasiparticles is exponentially suppressed by the gap for $T\ll\Delta$.
Thus, the screening current is given by Eq.~\eqref{eq-Current_Nonlinear_Perturbative}, for any $\vA(\vr,t)$ in the limits $\hbar\omega\ll 2\Delta$ and $T\ll T_c$. The EM field can host a large number of photons in any mode, $N_{k}\gg 1$, provided the surface field is below the threshold for vortex generation. In this case we can fix the gauge and absorb the condensate phase into the vector potential: $\vA-\frac{\hbar c}{2e}\grad\vartheta\rightarrow\vA$. The current response in the superconductor in the transverse gauge can then be expressed as,
\be\label{eq-Current_Nonlinear_A}
\vj_s=-\frac{c}{4\pi\lambda_{\text{L}}^2}
\left\{1 - \frac{\theta}{A_c^2}\vert\vA\vert^2\right\}\,\vA
\,,
\ee
where 
$\lambda_{\text{L}}$ is the London penetration depth, $A_c\equiv H_c\lambda_{\text{L}}$, and $H_c$ is the critical field.

For photons on the vacuum side of the cavity the screening current reflects the radiation back into the vacuum. Thus, for photons in two resonant modes with frequencies $\omega_1$ and $\omega_2$ the superconductor provides nearly perfect lossless reflection of incident photons.
However, the cubic nonlinearity of the screening current of Eq.~\eqref{eq-Current_Nonlinear_A}, proportional to $|\vA|^2\,\vA$, generates current sources,
$\vj_s(\omega_a)$, at the vacuum-superconductor interface with frequencies,
$\omega_a\in\left\{\omega_1\,, \omega_2\,, 3\omega_1\,, 3\omega_2\,, 2\omega_1\pm\omega_2\,, 2\omega_2\pm\omega_1\right\}$,
that radiate photons into the vacuum at third harmonic and intermodulation frequencies distinct from those of the pump modes~\cite{sau22}.
If any of these frequencies correspond to resonant modes of the cavity, then the large mode density $\propto Q$ will lead to significant photon conversion from the pump modes to the resonant mode. We analyze this situation in detail since it is a potentially relevant background source of photons in axion searches and tests of QED based on photon-photon scattering in SRF cavities.

\subsection{Frequency Shift of Cavity Resonances by Nonlinear Meissner Currents}

Penetration of the EM field into the superconductor also leads to small changes in the resonant frequencies of modes compared to the geometrically determined frequencies of a perfect conductor. For Nb these frequency shifts are typically of order $\delta f\sim 10\,\mbox{kHz}$ for modes with $f\approx 1-10\,\mbox{GHz}$, and depend on the material properties of the superconductor that determine the London penetration depth, and thus temperature and frequency~\cite{uek22,zar23}. In the linear response limit the frequency shift of a mode is determined by the London penetration depth and is independent of the EM power (photon number) in the mode.
However, the nonlinear contribution to the Meissner screening current leads to an additional frequency shift that depends on the EM power (photon number). 
This power dependence of the frequency shift can, in principle, be used to differentiate signal photons from axion frequency conversion from photons generated by nonlinear Meissner screening currents.

In what follows we report calculations of the number of NLM photons, leakage of photons in the pump modes into the signal mode (hereafter ``leakage noise''), and the resulting impact of these sources on the sensitivity of SRF cavities to axion and QED mediated photon conversion. 
Our analysis also agrees with earlier calculations of the number of photons generated by the EH interaction in the volume of the cavity. 

For the single cavity setup proposed in Ref.~\cite{bog19} we show that for SRF cavities with ultra-high $Q\sim10^{22}$, the NLM effect parametrically shifts surface generated photons away from the signal frequency sufficiently to allow for detection of nonlinear QED conversion by the EH interaction. Detection also requires suppression of unbound electrons in the superconducting cavity, and thus operational temperatures of $T\approx 0.1$ K. 
%

In Sec.~\ref{sec:ii} we derive the EM fields at the surface of the cavity resulting from the NLM effect and leakage noise at the intermodulation frequency generated by two pump modes in a single SRF cavity.
Our analysis is based on Slater's method for calculating the EM fields in hollow cavities~\cite{sla46}.
In Sec.~\ref{sec:iii} we report our results for the number of photons due to the NLM effect and leakage noise, and compare them with predictions based on the EH interaction.
Finally, we discuss the impact of the NLM effect and leakage noise on sensitivity of the single cavity setup to the axion signal, and related advantages of the emitter/receive cavity setup proposed by Gao and Harnik~\cite{gao21}.

\section{Formalism: Slater's method\label{sec:ii}} 

We consider a cylindrical SRF cavity pumped simultaneously with two resonant frequencies $\omega_1$ and $\omega_2$, and calculate the NLM signal and leakage noise fields from the pump fields. The axion field, EH term, and NLM effect give signal photons at intermodulation frequency $\omega_3=2\omega_1-\omega_2$ due to their cubic nonlinearity. 
We choose solenoidal vector functions at three resonant mode frequencies: $\omega_1^\infty=\omega_{011}^{\rm TE}$, $\omega_2^\infty=\omega_{010}^{\rm TM}$, and $\omega_3^\infty=\omega_{020}^{\rm TM}$ for a cylindrical cavity of radius $R$ and height $L=3.112R$, where the superscript $\infty$ denotes the ideal ``perfect conducting'' cavity mode frequenciess~\cite{bog19}.
The corresponding TM and TE mode frequencies are given by $\omega_{nml}^{\rm TM}=c\sqrt{(Z_{nm}/R)^2+(l\pi/L)^2}$ and $\omega_{nml}^{\rm TE}=c\sqrt{(S_{nm}/R)^2+(l\pi/L)^2}$, with $Z_{nm}$ and $S_{nm}$ denoting the $m$th zero and extremum of the $n$th Bessel function of the first kind $J_n(\rho)$, respectively (See App. \ref{app:a}). One can check that the TM$_{020}$ mode corresponds to the desired intermodulation frequency: $\omega_{020}^{\rm TM}=2\omega_{011}^{\rm TE}-\omega_{010}^{\rm TM}$ using $Z_{01}=2.405$, $S_{01}=Z_{11}=3.832$, and $Z_{02}=5.521$~\cite{hill09}. 
The exact mode frequencies include the frequency shifts relative to $\omega_i^\infty$ resulting from field pentration into the superconductor as discussed in Sec. \ref{sec:iid}.

Our formalism for predicting photon conversion within SRF cavities follows Slater~\cite{sla46}. We expand the EM field in the electric and magnetic basis functions of an ideal cavity for each mode, i.e. ${\bf e}_i({\bf r})$ and ${\bf b}_i({\bf r})$ for $i\in\{1,2,3\}$.
We then obtain the equation for the amplitude of the electric field, ${\bf E}_3({\bf r},t)={\bf e}_3({\bf r})\int_V{\bf E}({\bf r}',t)\cdot{\bf e}_3({\bf r}')dv'$, at the intermodulation mode frequency, $\omega_3$, as (See App. \ref{app:b} for details),
\begin{equation}\label{eq-cavityE-2}
\left(\frac{1}{c^2}\frac{d^2}{dt^2}
+
\frac{\omega_3^\infty{}^2}{c^2}\right)
\int_V{\bf E}({\bf r},t)\cdot{\bf e}_3({\bf r})dv 
=
-\frac{4\pi}{c^2}\frac{d}{dt}\int_V{\bf j}({\bf r},t)\cdot{\bf e}_3({\bf r})dv
-\frac{\omega_3^\infty}{c}
\int_S[{\bf E}({\bf r},t)\times{\bf b}_3({\bf r})]\cdot d{\bf a}
\,. 
\end{equation}
The first term on the right-hand side of Eq.~\eqref{eq-cavityE-2} is the vacuum term that includes source currents from the axion field and the EH interaction~\cite{bog19}. 
The second term is the conductive surface term, which contains the NLM source currents as well as the leakage of the pump modes into the signal mode.
Bogorad \emph{et al.} considered only the vacuum term and calculated the signal EM fields from the effective charge currents generated by the axion field and EH interaction~\cite{bog19}, 
\begin{eqnarray} 
\frac{4\pi}{c}{\bf j}_a({\bf r},t)
&=&
g_{a\gamma\gamma}\,
\left[
{\bf E}({\bf r},t)\times{\bm\nabla}a({\bf r},t)
-{\bf B}({\bf r},t)\frac{1}{c}\frac{\partial a({\bf r},t)}{\partial t}
\right]
\,,
\label{eq-axion_current}
\end{eqnarray} 
where the axion field $a({\bf r},t)$ is sourced by the pseudo-scalar invariant of the EM field~\cite{sik83,bog19},
\begin{equation} 
\left(
\frac{1}{c^2}\partial_t^2 - {\bm\nabla}^2 +\frac{m_a^2c^2}{\hbar^2}
\right)\,
a({\bf r},t) = g_{a\gamma\gamma}{\bf E}({\bf r},t)\cdot{\bf B}({\bf r},t)
\,.
\label{eq-axion_field}
\end{equation}
Similarly, the corresponding EH current density is given by  
\begin{eqnarray} 
\frac{4\pi}{c}{\bf j}_{\rm EH}({\bf r},t)
=
\frac{\hbar e^4}{45\pi m_e^4c^7}
\Bigg[
\ns\ns&&\ns\ns
\frac{1}{c}\frac{\partial}{\partial t}
\Big\{
2[E^2({\bf r},t)-B^2({\bf r},t)]\,{\bf E}({\bf r},t) 
+7[{\bf E}({\bf r},t)\cdot{\bf B}({\bf r},t)]
{\bf B}({\bf r},t)
\Big\}
\nonumber\\
&-&
{\bm\nabla}\times
\Big\{
2[
E^2({\bf r},t)-B^2({\bf r},t)
]{\bf B}({\bf r},t) 
-7[{\bf E}({\bf r},t)\cdot{\bf B}({\bf r},t)]
{\bf E}({\bf r},t)
\Big\}
\Bigg]
\,,
\label{eq-EH_current}
\end{eqnarray} 
From these equations we can calculate the number of photons converted to the signal mode by the EH interaction based on Eq.~\eqref{eq-EH_current}, and sensitivity to the axion-photon coupling and axion mass based on Eqs.~\eqref{eq-axion_current} and~\eqref{eq-axion_field}. Our analysis follows closely, and the results agree with, that reported by Bogorad et al.~\cite{bog19}.
In the next section we develop a method for calculating photon conversion via the NLM source currents at the intermodulation frequency, as well as the leakage noise from the surface term in Eq.~\eqref{eq-cavityE-2} based on the cavity perturbation method of Slater~\cite{sla46} combined with the theory of the nonlinear Meissner currents for superconducting resonators~\cite{sau22}.

\subsection{NLM Signal and Leakage Fields\label{sec:iib}}

Here we calculate the EM field generated by the NLM screening current at the signal frequency, as well as the field resulting from the leakage of the pump modes into the signal mode~\cite{sau22}.
We start from an electric field that is a sum of the fields of the pump modes and the resonant intermodulation (signal) mode,
\begin{equation}\label{eq-form_E}
{\bf E}({\bf r},t)=\sum_{i=1}^3{\bf E}_i({\bf r},t)
\,,\quad\mbox{with}\quad
{\bf E}_i({\bf r},t)={\rm Re}{\bf E}_i({\bf r}){\rm e}^{-i\omega_it-\frac{\omega_i^\infty}{2Q_i}t}
\,, 
\end{equation} 
where $Q_i$ is the quality factor of the $i^{\mbox{th}}$ mode at frequency $\omega_i=\omega_i^\infty+\delta\omega_i$, which includes the frequency shift $\delta\omega_i$ resulting from the penetration of the field ${\bf E}_i({\bf r},t)$ into the superconductor~\cite{jackson67,padamsee08,gur17}. For high-$Q$ resonances we have $Q_i\gg1$ for all three modes. 
Note that Bogorad \emph{et al}. neglected the frequency shift in their analysis~\cite{bog19}. In the case of photon conversion via the NLM currents the frequency shifts for pump and signal modes are inequivalent, and this fact is relevent to the number of signal photons generated via both the NLM effect and leakage into the signal mode.

Substituting Eq.~\eqref{eq-form_E} into the surface term of Eq.~\eqref{eq-cavityE-2}, multiplying both sides by ${\rm e}^{i\omega_3t-\frac{\omega_3^\infty}{2Q_3}t}$, then averaging over time we obtain the 
complex amplitude and spatial profile of the electric field of the signal mode,
\begin{align} 
{\bf E}_3({\bf r})
&=
i\sum_{i=1}^3C_i\frac{c\,Q_3}{\omega_3^\infty}{\bf e}_3({\bf r})
\int_S[{\bf E}_i({\bf r}')\times{\bf b}_3({\bf r}')]\cdot d{\bf a}', \label{eq-E3}
\end{align}
where $C_i$ is given by 
\begin{equation} 
C_i\equiv 
C_3\,
\frac{-i\omega_3^\infty/Q_3}{1-{\rm e}^{-\frac{\omega_3^\infty}{Q_3}t_p}}
\frac{1-{\rm e}^{-i(\omega_i-\omega_3)t_p-\frac{\omega_i^\infty}{2Q_i}t_p-\frac{\omega_3^\infty}{2Q_3}t_p}}{\omega_i-\omega_3-i\omega_i^\infty/2Q_i-i\omega_3^\infty/2Q_3}
=
C_3\,
\left\{
\begin{array}{cll} 
\displaystyle 1
&,
&\omega_i^\infty t_p\ll1,
\\
\displaystyle
\frac{1-{\rm e}^{-i(\omega_i-\omega_3)t_p}}{-i(\omega_i-\omega_3)t_p} 
&,
&1\ll\omega_i^\infty t_p\ll Q_i, 
\\
\displaystyle
\frac{(-i\omega_3^\infty/Q_3)}{\omega_i-\omega_3-i\omega_i^\infty/2Q_i-i\omega_3^\infty/2Q_3} 
&,
&Q_i\ll\omega_i^\infty t_p,
\end{array}\right.
\label{eq-Ci}
\end{equation} 
$t_p$ is the period of the pump field, and
\begin{equation}\label{eq-C3}
C_3
\equiv
\frac{-i\omega_3^\infty{}^2/Q_3}{\omega_3^2-\omega_3^\infty{}^2-i\omega_3\omega_3^\infty/Q_3}
\approx
\frac{-i\omega_3^\infty/2Q_3}{\omega_3-\omega_3^\infty-i\omega_3^\infty/2Q_3}
\,.
\end{equation}

For the TE$_{011}$ (mode 1) and TM$_{010}$ (mode 2) pump modes, and TM$_{020}$ (mode 3) signal mode 
we have ${\bf E}_1({\bf r})=E_{1\varphi}(\rho,z)\,\hat{\bm\varphi}$,
${\bf E}_2({\bf r})=E_{2z}(\rho)\,\hat{\bf z}$, 
${\bf E}_3({\bf r})=E_{3z}(\rho)\,\hat{\bf z}$, 
and ${\bf b}_3({\bf r})=b_{3\varphi}(\rho)\,\hat{\bm\varphi}$. 
Thus, the contribution to Eq.~\eqref{eq-E3} from ${\bf E}_1({\bf r})$ vanishes. The result for Eq.~\eqref{eq-E3} can then be written as 
\begin{subequations} 
\begin{align} 
{\bf E}_3({\bf r})&={\bf E}_{\rm leak}({\bf r})+{\bf E}_{\rm NLM}({\bf r}), 
\label{eq-NLM_leak_E_sum}
\\
{\bf E}_{\rm leak}({\bf r})
&=-2\pi iRLC_2\frac{c\,Q_3}{\omega_3^\infty}E_{2z}(R)\,b_{3\varphi}(R)\,{\bf e}_3({\bf r})
\,, 
\\
{\bf E}_{\rm NLM}({\bf r})
&=-2\pi iRLC_3\frac{c\,Q_3}{\omega_3^\infty}E_{3z}(R)\,b_{3\varphi}(R)\,{\bf e}_3({\bf r})
\,, 
\end{align}
\label{eq-NLM_leak_E}
\end{subequations} 
where ${\bf E}_{\rm leak}({\bf r})$ is the field resulting from the leakage of the pump fields into the signal mode, and ${\bf E}_{\rm NLM}({\bf r})$ is the field of the signal mode due to the NLM effect.

\begin{figure*}[t]
\begin{center}
\includegraphics[width=0.8\linewidth]{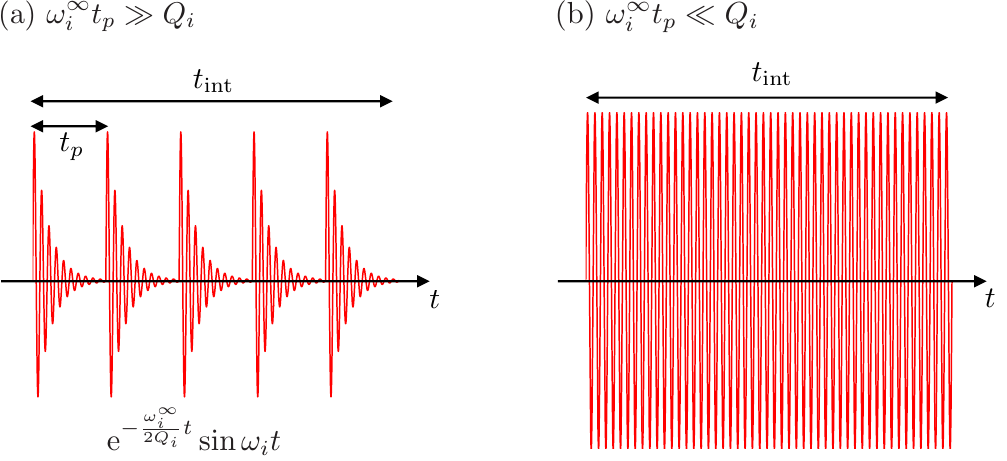}
\caption{
A schematic illustrating the pumping period and experimental integration (total measurement) times, $t_p$ and $t_{\rm int}$, respectively, for the time dependent pump field, $E_{pi}(t)={\rm e}^{-\frac{\omega_i^\infty}{2Q_i}t}\sin\omega_i t$. (a) For $\omega_i^\infty t_p\gg Q_i$ the damping of the pump field is significant during each pump cycle. (b) For $\omega_i^\infty t_p\ll Q_i$ the damping rate is $\sim 1/t_{\rm int}$.
}
\label{fig-pumping_periodic_time}
\end{center}
\end{figure*}

Figure~\ref{fig-pumping_periodic_time} is a representation of the period of pump fields, $t_p$, and the total integration time, $t_{\rm int}$, for a measurement of photon conversion to the signal mode.
If $\omega_i^\infty t_p\ll Q_i$, the pump fields are undamped until the measurement is stopped. We can then replace the cavity bandwidths, $B_i=\omega_i^\infty/Q_i$, with $B_i=1/t_{\rm int}$, where $t_{\rm int}$ is the integration time for measurement of photon conversion into the signal mode~\cite{bog19,ber20,gao21}. In this limit the signal power is large, but the leakage noise may be also significant since $C_2=[1-{\rm e}^{-i(\omega_2-\omega_3)t_p}]C_3/[-i(\omega_2-\omega_3)t_p]$. 
However, if $\omega_i^\infty t_p\gg Q_i$ the leakage noise is greatly suppressed due to the frequency difference of the order of GHz between the pump and signal modes.

\subsection{Field Amplitudes on the Cavity Walls\label{sec:iic}} 

The frequency shifts of the pump and signal modes can be calculated by first computing the amplitude of electric field on the cavity walls, neglecting the damping of the modes (See Sec. \ref{sec:iid}).
The surface integral for a cylindrical cavity at the radial wall, $\rho=R$, is carried out for the NLM screening profile. Since the cavity radius is much larger than the penetration depth, $R\gg\lambda_{\rm L}$ we can neglect the curvature of the cylindrical cavity wall.
Then, using Faraday's law (Eq.~\eqref{eq-Faraday-2}), ${\bf E}_i({\bf r})=E_{iz}({\rho})\hat{\bf z}$, and ${\bf B}_i({\bf r})=B_{i\varphi}({\rho})\hat{\bm\varphi}$ for $i=2,3$, we obtain the amplitude of electric field on the cavity wall
\begin{align} 
E_{iz}(R)=-\frac{i\omega_i^\infty}{c}\int_\infty^RdxB_{i\varphi}(x)
\,.
\label{eq-NLM_leak_surface_E}
\end{align}
The penetration of the magnetic field into the superconductor is given by Eq.~(91) of Ref.~\cite{sau22} in the limit $\hbar\omega_i\ll 2\Delta$, and was obtained by solving the nonlinear London equation (c.f. App.~\ref{app:c}). The resulting spatial integral of the magnetic field on the right-hand side of Eq.~\eqref{eq-NLM_leak_surface_E} is then, 
\begin{equation}
\int_\infty^R dx B_{i\varphi}(x,t) 
=\int_\infty^0 dx'
\frac{H{\rm e}^{-x'/\lambda_{\rm L}}}
{[1+\theta(H/2H_c)^2({\rm e}^{-2x'/\lambda_{\rm L}}-1)]^{3/2}} 
\approx
-\lambda_{\rm L}H
\left[1+\frac{\theta}{4}\left(\frac{H}{H_c}\right)^2\right]
\,,
\label{eq-integral_B_SC}
\end{equation}
where $H$ is the magnetic field at the surface and the right side of Eq.~\eqref{eq-integral_B_SC} is the approximate result to leading order in the nonlinear field dependence. The field scale is set by the thermodynamic critical field $H_c\equiv(c/e)(\Delta/v_f\lambda_{\rm L})$ where $e$ and $v_f$ are the electron charge and Fermi velocity, respectively. The other material parameters are the London penetration depth, $\lambda_{\rm L}$, and $\theta=\theta(T;\tau)$ is the dimensionless coefficient for the nonlinear correction to the Meissner screening current, and is given by Eq.~(85) in Ref.~\cite{sau22} with $T$ and $\tau$ denoting temperature and mean electron-impurity scattering time, respectively. 

The nonlinear contribution to the surface magnetic field at the intermodulation frequency is determined by the magnetic fields of the pump modes,
\begin{align} 
{\bf B}_p({\bf r},t)&={\bf B}_1^\infty({\bf r},t)+{\bf B}_2^\infty({\bf r},t), 
\end{align}
where ${\bf B}_1^\infty({\bf r},t)$ is the TE$_{011}$ mode and ${\bf B}_2^\infty({\bf r},t)$ is the TM$_{010}$ mode (App.~\ref{app:a}), 
\begin{subequations} 
\begin{align} 
{\bf B}_1^\infty({\bf r},t)
&=\left[B_{1\rho}^\infty({\bf r})\hat{\bm\rho}+B_{1z}^\infty({\bf r})\hat{\bf z}
\right]\cos\omega_1^\infty t, \\
{\bf B}_2^\infty({\bf r},t)
&=-iB_{2\varphi}^\infty({\bf r})\hat{\bm\varphi}\sin\omega_2^\infty t. 
\end{align}
\end{subequations} 
The resulting cubic contribution to the surface field reduces to
\begin{equation}\label{eq-H^3}
H^3\approx i B_{2\varphi}^{\infty3}(R)\sin^3\omega_2^\infty\,t 
-\frac{i}{2}B_{1z}^{\infty2}(R)B_{2\varphi}^\infty(R)\cos^2\omega_1^\infty t\sin\omega_2^\infty\,t
\,, 
\end{equation}
after averaging the $z$-dependence of $|B_{1z}^\infty(\rho,z)|^2\propto\sin(\pi z/L)$ over the length of the cavity and using the fact that $B_{1\rho}(R)\equiv 0$ (See App.~\ref{app:a}). Then, using $\cos^2\omega_1^\infty t\sin\omega_2^\infty t =-(1/4)\sin\omega_3^\infty t +(1/4)\sin((2\omega_1^\infty+\omega_2^\infty)t) +(1/2)\sin\omega_2^\infty t$ we obtain the surface electric fields of the pump and intermodulation modes,
\begin{subequations} 
\begin{align} 
E_{2z}(R)
=&
+\frac{i\omega_2^\infty}{c}\lambda_{\rm L}
\times
\left[1+\frac{\theta}{32H_c^2}|B_{2\varphi}^\infty(R)|^2\right]B_{2\varphi}^\infty(R), \\
E_{3z}(R)
=&
-\frac{i\omega_3^\infty}{c}\lambda_{\rm L}
\times
\frac{\theta}{32H_c^2}|B_{1z}^\infty(R)|^2B_{2\varphi}^\infty(R)
\,.
\end{align}
\label{eq-NLM_leak_surface_E-2}%
\end{subequations} 

\subsection{Surface Impedance and Frequency shifts\label{sec:iid}}

We tacitly assumed in Sec.~\ref{sec:iic} that photon conversion via the NLM current and the leakage noise occur at the same frequency, i.e. $\omega_3^\infty$. However, as we show below the photon frequency from these two sources depends on the amplitudes of and NLM source field and the leakage noise field, and thus NLM photons and leakage noise photons are generated at slightly different frequencies.
Thus, to account for the nonlinear frequency shifts we write Eqs.~\eqref{eq-Ci} and~\eqref{eq-NLM_leak_E}, as 
\begin{subequations} 
\begin{align} 
{\bf E}_{\rm leak}({\bf r})
&=-2\pi iRLC_{\rm leak}\frac{cQ_3}{\omega_3^\infty}E_{2z}(R)b_{3\varphi}(R){\bf e}_3({\bf r}), \\
{\bf E}_{\rm NLM}({\bf r})
&=-2\pi iRLC_{\rm NLM}\frac{cQ_3}{\omega_3^\infty}E_{3z}(R)b_{3\varphi}(R){\bf e}_3({\bf r}), 
\end{align}\label{eq-NLM_leak_E-2}%
\end{subequations} 
where
\begin{subequations} 
\begin{align} 
C_{\rm leak}
&\equiv
\frac{-i\omega_3^\infty/Q_3}{1-{\rm e}^{-\frac{\omega_3^\infty}{Q_3}t_p}}
\times
\frac{-i\omega_3^\infty/2Q_3}{\omega_{\rm leak}-\omega_3^\infty-i\omega_3^\infty/2Q_3} 
\times
\frac{1-{\rm e}^{-i(\omega_2-\omega_{\rm leak})t_p-\frac{\omega_2^\infty}{2Q_2}t_p-\frac{\omega_3^\infty}{2Q_3}t_p}}{\omega_2-\omega_{\rm leak}-i\omega_2^\infty/2Q_2-i\omega_3^\infty/2Q_3}
\,,
\\
C_{\rm NLM}
&\equiv
\frac{-i\omega_3^\infty/2Q_3}{\omega_{\rm NLM}-\omega_3^\infty-i\omega_3^\infty/2Q_3}, 
\end{align}
\end{subequations} 
where
\begin{subequations} 
\begin{align} 
\omega_{\rm leak} 
&=\omega_3^\infty+\delta\omega_{\rm leak}(B_{\rm leak}(R)), \\
\omega_{\rm NLM} 
&=\omega_3^\infty+\delta\omega_{\rm NLM}(B_{\rm NLM}(R)), 
\end{align}
\end{subequations} 
where $B_{\rm leak}(R)$ and $B_{\rm NLM}(R)$ are the amplitudes of surface magnetic fields resulting from the leakage of the pump modes and the NLM current, respectively. 
These fields are obtained from the equation for the magnetic field at the intermodulation 
frequency, $\omega_3^\infty$,
\begin{equation} 
\left(\frac{1}{c^2}\frac{d^2}{dt^2}+\frac{\omega_3^\infty{}^2}{c^2}\right)
\int{\bf B}({\bf r},t)\cdot{\bf b}_3({\bf r})dv
=
-\frac{1}{c}\frac{d}{dt}
\int_S[{\bf E}({\bf r},t)\times{\bf b}_3({\bf r})]\cdot d{\bf a}
\,, 
\end{equation}
and thus the surface magnetic fields,
\begin{subequations} 
\begin{align} 
B_{{\rm leak}\varphi}(R)
&=-2\pi RLC_{\rm leak}\frac{\omega_2^\infty cQ_3}{\omega_3^{\infty2}}E_{2z}(R)\,b_{3\varphi}^2(R)
\,,
\\
B_{{\rm NLM}\varphi}(R)
&=-2\pi RLC_{\rm NLM}\frac{cQ_3}{\omega_3^\infty}E_{3z}(R)\,b_{3\varphi}^2(R)
\,.
\end{align}
\end{subequations} 
Self-consistent equations for the frequency shift $\delta\omega_{\rm NLM}$ are 
obtained by first expressing the NLM fields as 
\begin{subequations}
\begin{align} 
{\bf E}_{\rm NLM}({\bf r},t)
&={\rm Re}{\bf E}_{\rm NLM}({\bf r})
{\rm e}^{-i(\omega_3^\infty+\delta\omega_{\rm NLM})t-\frac{\omega_3^\infty}{2Q_3}t}
\,,
\\
{\bf B}_{\rm NLM}({\bf r},t)
&={\rm Re}{\bf B}_{\rm NLM}({\bf r})
{\rm e}^{-i(\omega_3^\infty+\delta\omega_{\rm NLM})t-\frac{\omega_3^\infty}{2Q_3}t}
\,, 
\end{align}
\label{eq-form_EM_NLM}
\end{subequations}
then using Eqs.~\eqref{eq-form_EM_NLM} and Slater's method~\cite{sla46,uek22}, 
to express the frequency shift as
\begin{align} 
\delta\omega_{\rm NLM}=-\frac{\omega_3^\infty}{2G_3}X_{\rm NLM}^s
\,,
\label{eq-delta_omega_NLM}
\end{align}
where $X_{\rm NLM}^s$ is the NLM contribution to the surface reactance,
\begin{align} 
G_3\equiv\frac{Z_0}{\displaystyle\frac{c}{\omega_3^\infty}\int_S|{\bf b}_3({\bf r})|^2da}, 
\label{eq-G}
\end{align}
is the geometric factor of the signal mode, and $Z_0=4\pi/c=376.7\,{\rm\Omega}$ is the impedance of the vacuum.

The EM fields generated by the NLM currents are those of the TM$_{020}$ mode, and have the form ${\bf E}_{\rm NLM}({\bf r})=E_{{\rm NLM}z}(\rho)\hat{\bf z}$ and ${\bf B}_{\rm NLM}({\bf r})=B_{{\rm NLM}\varphi}(\rho)\hat{\bm\varphi}$. The corresponding surface reactance for this mode can be obtained from Faraday's law [Eq.~\eqref{eq-Faraday-2}] as shown below. As in Sec.~\ref{sec:iib} we can neglect the curvature of 
the cavity wall, in which case the surface magnetic field is given by
\begin{equation} 
H={\rm Re}B_{{\rm NLM}\varphi}(R){\rm e}^{-\frac{\omega_3^\infty}{2Q_3}t}\cos\omega_{\rm NLM}t
+{\rm Im}B_{{\rm NLM}\varphi}(R){\rm e}^{-\frac{\omega_3^\infty}{2Q_3}t}\sin\omega_{\rm NLM}t, \label{eq-H-NLM}
\end{equation}
The time averaged electric field $E_{{\rm NLM}z}(R)$ at the surface reduces to
\begin{equation}
E_{{\rm NLM}z}(R) 
=i\frac{\omega_3^\infty}{c}\lambda_{\rm L}B_{{\rm NLM}\varphi}(R)
\times\left[
1+\frac{\theta}{16H_c^2}
\frac{1-{\rm e}^{-\frac{3\omega_3^\infty}{2Q_3}t_p}}{1-{\rm e}^{-\frac{\omega_3^\infty}{2Q_3}t_p}}
|B_{{\rm NLM}\varphi}(R)|^2
\right]
\,, 
\label{eq-E_signal-2} 
\end{equation}
and the corresponding surface impedance, $Z_{\rm NLM}^s=R_{\rm NLM}^s-iX_{\rm NLM}^s$, is given by 
\begin{align}\label{eq-Zs-2}
Z_{\rm NLM}^s=-Z_0\frac{E_{{\rm NLM}z}(R)}{B_{{\rm NLM}\varphi}(R)}
\,.
\end{align}
Thus, we obtain the surface reactance at low temperatures from 
Eqs.~\eqref{eq-E_signal-2} and \eqref{eq-Zs-2},
\begin{align} 
X_{\rm NLM}^s
=Z_0\frac{\omega_3^\infty}{c}\lambda_{\rm L}
\times
\left[
1+
\frac{\theta}{16H_c^2}
\frac{1-{\rm e}^{-\frac{3\omega_3^\infty}{2Q_3}t_p}}{1-{\rm e}^{-\frac{\omega_3^\infty}{2Q_3}t_p}}
|B_{{\rm NLM}\varphi}(R)|^2
\right]. \label{eq-Xs}
\end{align}
The field independent term in Eq.~\eqref{eq-Xs} is the same as the surface reactance at low temperatures given by Eq.~(E3) of Ref.~\cite{gur17}.
However, the frequency shift also includes the contribution from the field-dependent NLM effect obtained from Eqs.~\eqref{eq-delta_omega_NLM} and \eqref{eq-Xs},
\begin{equation}\label{eq-nonlinear_delta_omega_NLM}
\delta\omega_{\rm NLM}=-\frac{\omega_3^{\infty2}\lambda_{\rm L}Z_0}{2cG_3} 
\times\left[
1+
\frac{\theta}{16H_c^2}
\frac{1-{\rm e}^{-\frac{3\omega_3^\infty}{2Q_3}t_p}}{1-{\rm e}^{-\frac{\omega_3^\infty}{2Q_3}t_p}}
|B_{{\rm NLM}\varphi}(R)|^2
\right]
\,.
\end{equation}
Similarly, the frequency shift of photons from the leakage of the pump modes into the intermodulation
mode is given by 
\begin{align}\label{eq-nonlinear_delta_omega}
\delta\omega_{\rm leak}=-\frac{\omega_2^\infty\omega_3^\infty\lambda_{\rm L}Z_0}{2cG_3}
\times\left[
1+
\frac{\theta}{16H_c^2}
\frac{1-{\rm e}^{-\frac{3\omega_3^\infty}{2Q_3}t_p}}{1-{\rm e}^{-\frac{\omega_3^\infty}{2Q_3}t_p}}
|B_{{\rm leak}\varphi}(R)|^2
\right]
\,.
\end{align}

The resonant frequencies of the pump modes are also shifted due to field penetration and damping from a cavity wall. Expressing the EM fields of the two pump modes ($i=1,2$),
\begin{subequations}
\begin{align} 
{\bf E}_i({\bf r},t)
&={\rm Re}{\bf E}_i({\bf r})
{\rm e}^{-i(\omega_i^\infty+\delta\omega_i)t-\frac{\omega_i^\infty}{2Q_i}t}
\,,
\\
{\bf B}_i({\bf r},t)
&={\rm Re}{\bf B}_i({\bf r})
{\rm e}^{-i(\omega_i^\infty+\delta\omega_i)t-\frac{\omega_i^\infty}{2Q_i}t}\
\,, 
\end{align}
\label{eq-form_EM_pump}
\end{subequations}
the resonant frequencies are,
\begin{subequations} 
\begin{align} 
\omega_1 
=\omega_1^\infty+\delta\omega_1(B_{1z}(R))
\,, 
\\
\omega_2 
=\omega_2^\infty+\delta\omega_2(B_{2\varphi}(R))
\,. 
\end{align}
\end{subequations} 
Following the same analysis procedure as described for the calculation of the frequency shift of photons generaged by the NLM effect we obtain the frequency shifts $\delta\omega_1$ and $\delta\omega_2$ for 
photons in the pump modes,
\begin{subequations} 
\begin{align} 
\label{eq-delta_omega_1} 
\delta\omega_1=
-\frac{\omega_1^{\infty2}\lambda_{\rm L}Z_0}{\pi cG_1}
\times\left[
1+
\frac{\theta}{24H_c^2}
\frac{1-{\rm e}^{-\frac{3\omega_1^\infty}{2Q_1}t_p}}{1-{\rm e}^{-\frac{\omega_1^\infty}{2Q_1}t_p}}
|B_{1z}(R)|^2
\right],
\\
\label{eq-delta_omega_2}
\delta\omega_2=
-\frac{\omega_2^{\infty2}\lambda_{\rm L}Z_0}{2cG_2}
\times\left[
1+
\frac{\theta}{16H_c^2}
\frac{1-{\rm e}^{-\frac{3\omega_2^\infty}{2Q_2}t_p}}{1-{\rm e}^{-\frac{\omega_2^\infty}{2Q_2}t_p}}
|B_{2\varphi}(R)|^2
\right],
\end{align}
\end{subequations} 
where $B_{1z}(R)$ and $B_{2\varphi}(R)$ are obtained by replacing $\omega_i^\infty$ with $\omega_i$ in $B_{1z}^\infty(R)$ and $B_{2\varphi}^\infty(R)$. 
From the above equations for the resonant frequencies of photons in the two pump modes $\omega_1$ and $\omega_2$, the frequency of the signal photons sourced in the vacuum region of the cavity by the axion field or the EH interaction are given precisely by $\omega_s\equiv 2\omega_1-\omega_2$, i.e. only the nonlinear shifts of the two pump modes enter the intermodulation frequency for photons sourced by the axion field or EH interaction.

The NLM and leakage noise fields evaluated at the signal frequency, $\omega_s$, are given by Eq.~\eqref{eq-NLM_leak_E-2} with $\omega_3^\infty$ replaced by $\omega_s$, except in the equations for $\omega_{\rm leak}$ and $\omega_{\rm NLM}$,
\begin{subequations} 
\begin{align}
{\bf E}_{\rm leak}({\bf r})
&=
-2\pi iRLC_{\rm leak}\frac{cQ_s}{\omega_s}E_{2z}(R)\,
b_{s\varphi}(R){\bf e}_s({\bf r}),
\\
{\bf E}_{\rm NLM}({\bf r})
&=
-2\pi iRLC_{\rm NLM}\frac{cQ_s}{\omega_s}E_{3z}(R)\,
b_{s\varphi}(R){\bf e}_s({\bf r}), 
\end{align}
\end{subequations} 
where
\begin{subequations} 
\begin{align}
C_{\rm leak}
&=
\frac{-i\omega_s/Q_s}{1-{\rm e}^{-\frac{\omega_s}{Q_s}t_p}}
\times
\frac{-i\omega_s/2Q_s}{\omega_{\rm leak}-\omega_s-i\omega_s/2Q_s} 
\times
\frac{1-{\rm e}^{-i(\omega_2-\omega_{\rm leak})t_p-\frac{\omega_2}{2Q_2}t_p-\frac{\omega_s}{2Q_s}t_p}}{\omega_2-\omega_{\rm leak}-i\omega_2/2Q_2-i\omega_s/2Q_s}, \label{eq-C_leak} 
\\
C_{\rm NLM}
&=
\frac{-i\omega_s/2Q_s}{\omega_{\rm NLM}-\omega_s-i\omega_s/2Q_s}, \label{eq-C_NLM}
\end{align}
\end{subequations} 
and 
\begin{subequations}
\begin{align} 
B_{{\rm leak}\varphi}(R)
&=-2\pi RLC_{\rm leak}\frac{\omega_2cQ_s}{\omega_s^2}E_{2z}(R)b_{s\varphi}^2(R), \\
B_{{\rm NLM}\varphi}(R)
&=-2\pi RLC_{\rm NLM}\frac{cQ_s}{\omega_s}E_{3z}(R)b_{s\varphi}^2(R), \label{eq-surface_magnetic_field-NLM}
\end{align}
\label{eq-surface_magnetic_field}
\end{subequations} 
Note that $Q_s$ denotes the quality factor at $\omega_s$, while ${\bf e}_s({\bf r})$ and ${\bf b}_s({\bf r})$ are basis functions for the signal mode. 
They are equal to ${\bf e}_3({\bf r})$ and ${\bf b}_3({\bf r})$, respectively, to excellent approximation. The cavity damping effect is also included by extending Eq.~\eqref{eq-NLM_leak_surface_E-2},
\begin{subequations}
\begin{align} 
E_{2z}(R)
&=
\frac{i\omega_2^\infty}{c}\lambda_{\rm L}
\times \left[
1+\frac{\theta}{96H_c^2}
\frac{1-{\rm e}^{-\frac{3\omega_2^\infty}{2Q_2}t_p}}{1-{\rm e}^{-\frac{\omega_2^\infty}{2Q_2}t_p}}
|B_{2\varphi}(R)|^2
\right]B_{2\varphi}(R), 
\\
E_{3z}(R)
&=
-\frac{i\omega_3^\infty}{c}\lambda_{\rm L}\frac{\theta}{32H_c^2}
\frac{\omega_3^\infty t_p/2Q_3}{1-{\rm e}^{-\frac{\omega_3^\infty}{2Q_3}t_p}} 
\times
\frac{1-{\rm e}^{-\frac{\omega_1^\infty}{Q_1}t_p-\frac{\omega_2^\infty}{2Q_2}t_p}}{\omega_1^\infty t_p/Q_1+\omega_2^\infty t_p/2Q_2} 
|B_{1z}(R)|^2B_{2\varphi}(R).
\end{align}
\end{subequations} 
We can also use $B_{1z}(R)\approx B_{1z}^\infty(R)$ and $B_{2\varphi}(R)\approx B_{2\varphi}^\infty(R)$, which are an excellent approximation. By solving 
Eqs.~\eqref{eq-nonlinear_delta_omega_NLM},~\eqref{eq-nonlinear_delta_omega} and \eqref{eq-surface_magnetic_field} self-consistently, we can obtain the frequency shift as a function of the pump field strength.

\section{Results: NLM, leakage \& EH Photons\label{sec:iii}} 

The mean number of photons generated by the NLM current and the leakage of the pump field into the resonant signal frequency $\omega_s$ are given by
\begin{subequations}
\begin{align} 
N_{\rm NLM}
&=
\frac{1}{\hbar\omega_s}\int\frac{dv}{8\pi}|{\bf E}_{\rm NLM}({\bf r})|^2
\frac{1}{t_p}\int_0^{t_p}dt{\rm e}^{-\frac{\omega_3^\infty}{Q_s}t}
=
|C_{\rm NLM}|^2N_{\rm NLM}^{(0)}, 
\\
N_{\rm leak}
&=
\frac{1}{\hbar\omega_s}\int\frac{dv}{8\pi}|{\bf E}_{\rm leak}({\bf r})|^2
\frac{1}{t_p}\int_0^{t_p}dt{\rm e}^{-\frac{\omega_3^\infty}{Q_s}t}
=
|C_{\rm leak}|^2N_{\rm leak}^{(0)}, 
\end{align}
\label{eq-C_NLM+leak}
\end{subequations} 
where $N_{\rm NLM}^{(0)}$ and $N_{\rm leak}^{(0)}$ are the number of NLM and leakage photons that would be predicted if we neglected the nonlinear frequency shifts of the photons generated by the NLM and leakage fields,
\begin{subequations} 
\begin{align}
N_{\rm NLM}^{(0)}
&=
\frac{c^2Q_s^2VE_0^6}{8\pi\hbar\omega_s^3}
\frac{1-{\rm e}^{-\frac{\omega_3^\infty}{Q_s}t_p}}{\omega_3^\infty t_p/Q_s} 
\kappa_{\rm NLM}^2K_{\rm NLM}^2,
\\
N_{\rm leak}^{(0)}
&=
\frac{c^2Q_s^2VE_0^6}{8\pi\hbar\omega_s^3}\frac{\omega_s^2}{\omega_2^2}
\frac{1-{\rm e}^{-\frac{\omega_3^\infty}{Q_s}t_p}}{\omega_3^\infty t_p/Q_s}
\left(
\kappa_{\rm leak}^{{\rm L}2}K_{\rm leak}^{{\rm L}2} 
+
2\kappa_{\rm leak}^{\rm L}\kappa_{\rm leak}^{\rm NL}K_{\rm leak}^{\rm L}K_{\rm leak}^{\rm NL} 
+\kappa_{\rm leak}^{{\rm NL}2}K_{\rm leak}^{{\rm NL}2}
\right)
\,,
\end{align}
\end{subequations} 
where $V=\pi R^2L$ is the volume of the cavity, $E_0\equiv\sqrt{(1/V)\int dv|{\bf E}_1^\infty({\bf r})|^2}=\sqrt{(1/V)\int dv|{\bf E}_2^\infty({\bf r})|^2}$ is the spatially averaged electric field strength of either pump mode. The electric field of the pump modes are then ${\bf E}_1^\infty({\bf r})$ with $\omega_1^\infty=\omega_{011}^{\rm TE}$ and ${\bf E}_2^\infty({\bf r})$ with $\omega_2^\infty=\omega_{010}^{\rm TM}$, and given by (See App.~\ref{app:a}) 
\begin{subequations} 
\begin{align} 
&{\bf E}_1^\infty({\bf r})=
i\frac{\sqrt{2}E_0}{J_0(S_{01})} 
J_1\left(\frac{S_{01}}{R}\rho\right)\sin\frac{\pi z}{L}\hat{\bm\varphi}\,,
\\
&{\bf E}_2^\infty({\bf r})=
\frac{E_0}{J_1(Z_{01})}J_0\left(\frac{Z_{01}}{R}\rho\right)\hat{\bf z}
\,. 
\end{align}
\label{eq-pump_signal_E}%
\end{subequations} 
The number of resonant NLM and leakage photons are proportional to $Q_s^2$, $E_0^6$ and the factors $\kappa_{\rm NLM}$, $\kappa_{\rm leak}^{\rm L}$, $\kappa_{\rm leak}^{\rm NL}$, $K_{\rm NLM}$, $K_{\rm leak}^{\rm L}$, and $K_{\rm leak}^{\rm NL}$,
\begin{subequations} 
\begin{align} 
\kappa_{\rm NLM}
&\equiv
\frac{\omega_s^2}{c^2}
\frac{\omega_3^\infty t_p/2Q_s}{1-{\rm e}^{-\frac{\omega_3^\infty}{2Q_s}t_p}}
\frac{1-{\rm e}^{-\frac{\omega_1^\infty}{Q_1}t_p-\frac{\omega_2^\infty}{2Q_2}t_p}}
{\omega_1^\infty t_p/Q_1+\omega_2^\infty t_p/2Q_2} 
\frac{\lambda_{\rm L}\theta}{4H_c^2}
\,
\\
\kappa_{\rm leak}^{\rm L}
&\equiv
\frac{\omega_2^2}{c^2}\frac{\lambda_{\rm L}}{E_0^2}
\,,
\\
\kappa_{\rm leak}^{\rm NL}
&\equiv
\frac{\omega_2^2}{c^2}
\frac{1-{\rm e}^{-\frac{3\omega_2^\infty}{2Q_2}t_p}}{1-{\rm e}^{-\frac{\omega_2^\infty}{2Q_2}t_p}}
\frac{\lambda_{\rm L}\theta}{4H_c^2}
\,, 
\\
K_{\rm NLM}
&\equiv
\frac{c}{\omega_s\sqrt{V}}
\left|\int_S[\tilde{\bf E}_3({\bf r})\times{\bf b}_3({\bf r})]\cdot d{\bf a}\right|
\,, 
\\
K_{\rm leak}^{\rm L}
&\equiv
\frac{c}{\omega_s\sqrt{V}}
\left|\int_S[\tilde{\bf E}_2^{\rm L}({\bf r})\times{\bf b}_3({\bf r})]\cdot d{\bf a}\right|
\,, 
\\
K_{\rm leak}^{\rm NL}&\equiv\frac{c}{\omega_s\sqrt{V}}
\left|\int_S[\tilde{\bf E}_2^{\rm NL}({\bf r})\times{\bf b}_3({\bf r})]\cdot d{\bf a}\right|
\,,
\end{align}
\end{subequations} 
where ${\bf E}_3({\bf r}) \equiv(c/\omega_s)\kappa_{\rm NLM}E_0^3\tilde{\bf E}_3({\bf r})$, ${\bf E}_2^{\rm L}({\bf r})\equiv(c/\omega_2)\kappa_{\rm leak}^{\rm L}$ $E_0\tilde{\bf E}_2^{\rm L}({\bf r})$, and ${\bf E}_2^{\rm NL}({\bf r})\equiv(c/\omega_2)\kappa_{\rm leak}^{\rm NL}E_0^3\tilde{\bf E}_2^{\rm NL}({\bf r})$. The superscripts ``L" and ``NL" denote the linear and nonlinear terms, respectively. 

In the analysis to follow we choose the same cavity parameters used in Ref.~\cite{bog19}: $R=0.5$ m, $L=1.56$ m, and $f_3^\infty\equiv\omega_3^\infty/2\pi=527$ MHz, however we study the variation of photon conversion as a function of the pump field strength, quality factor, and cavity bandwidth. 
In what follows we adopt material parameters appropriate for N-doped Nb SRF cavities: the thermodynamic critical field $H_c=54$ MV/m, the zero-temperature energy gap $\Delta=1.55$ meV, the transition temperature $T_{c_0}=9.33$ K, the Fermi velocity $v_f=0.26\times10^8$ cm/s, the clean-limit London penetration depth $\lambda_{\rm L}=33$ nm~\cite{zar22}, and the dimensionless strength of the nonlinear Meissner current $\theta=0.001$~\cite{sau22}.
We also assume equal quality factors for the pump modes and the signal mode, i.e. $Q_1=Q_2=Q_s$.~\footnote{Thus, we neglect the weak frequency depdence, as well as the field-dependence, of the quality factor; the range of variation is relatively modest for $E_0$ below the maximum field gradient~\cite{pad17,gra13,rom14,padamsee08,gur17}}
The pump magnetic field profiles, ${\bf B}_1^\infty({\bf r})$ and ${\bf B}_2^\infty({\bf r})$, as well as the signal mode basis vector function, ${\bf b}_3({\bf r})$, are given by (See App.~ \ref{app:a}) 
\begin{eqnarray} 
{\bf B}_1^\infty({\bf r})
&=&
\frac{\sqrt{2}E_0}{J_0(S_{01})} 
\bigg[
-\frac{c}{\omega_1^\infty}J_1\left(\frac{S_{01}}{R}\rho\right) 
\frac{\pi}{L}\cos\frac{\pi z}{L}\hat{\bm\rho}
+\frac{c}{\omega_1^\infty}\frac{S_{01}}{R}J_0\left(\frac{S_{01}}{R}\rho\right)\sin\frac{\pi z}{L}\hat{\bf z}
\bigg], 
\label{eq-pump_M_1}
\\
{\bf B}_2^\infty({\bf r})
&=&
-i\frac{E_0}{J_1(Z_{01})}J_1\left(\frac{Z_{01}}{R}\rho\right)\hat{\bm\varphi}, 
\label{eq-pump_M_2} 
\\
{\bf b}_3({\bf r})
&=&
\frac{1}{\sqrt{V}J_1(Z_{02})}J_1\left(\frac{Z_{02}}{R}\rho\right)\hat{\bm\varphi}. \label{eq-signal_M_basis}
\end{eqnarray} 
These functions determine the coupling coefficients: $K_{\rm NLM}=0.084$, $K_{\rm leak}^{\rm L}=0.362$, and $K_{\rm leak}^{\rm NL}=0.271$. 

\begin{figure*}[t]
\begin{center}
\includegraphics[width=0.8\linewidth]{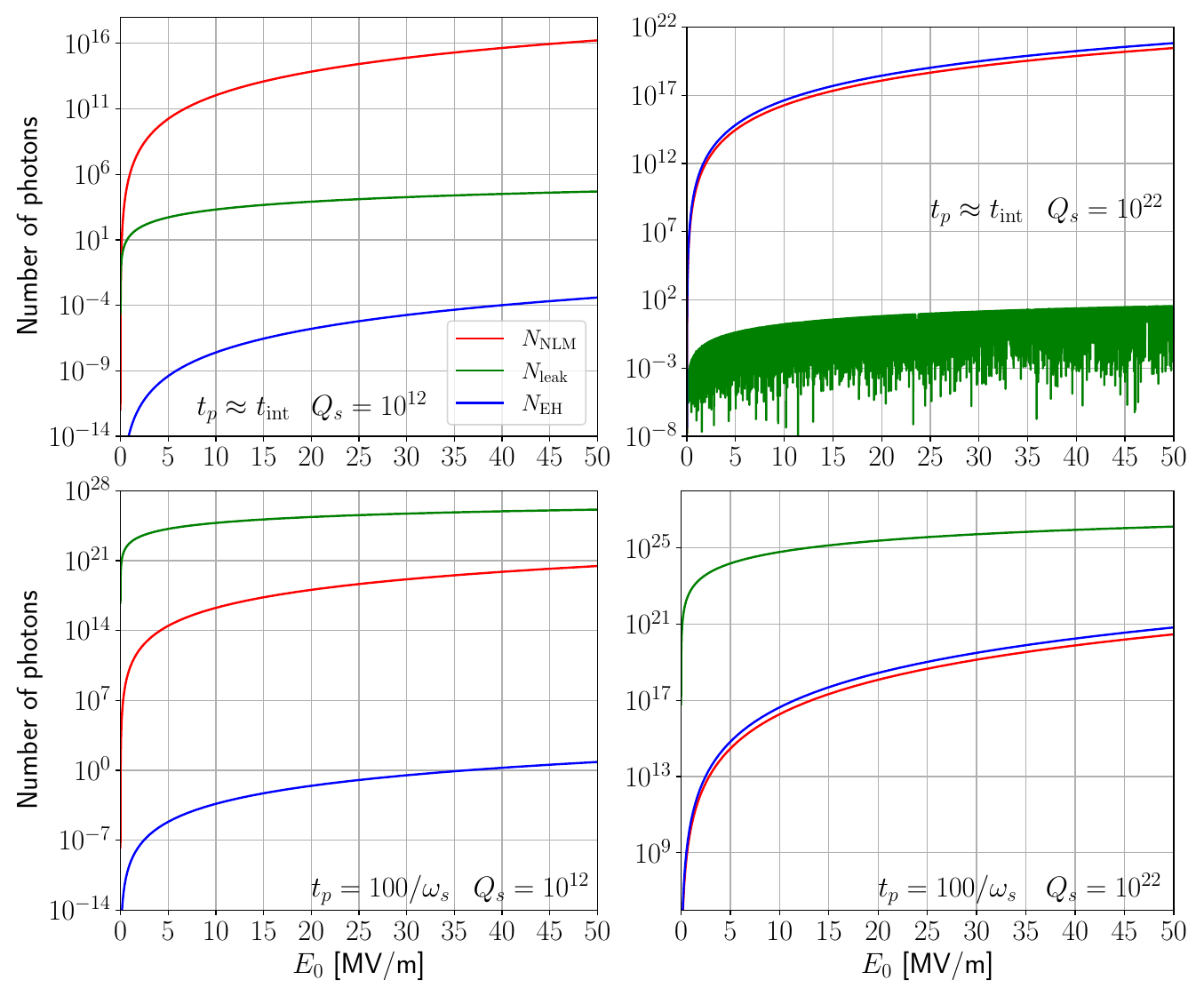}
\caption{Number of photons from NLM currents (red lines), leakage noise (green lines), and the EH interaction (blue lines) at the signal frequency $\omega_s$ as a function of the pump field strength $E_0$ for cavities with the quality factors, $Q_s=10^{12}$ (left panel) and $10^{22}$ (right panel), and pumping periods, $t_p=5\times10^{15}/\omega_s\approx t_{\rm int}$ (upper row) and $100/\omega_s$ (lower row).
The nonlinear coefficient used to calculate $N_{\rm NLM}$ is $\theta=0.001$.}
\label{fig-N}
\end{center}
\end{figure*}

The EH interaction provides a baseline for detection of rare events in the 2-pump mode, single-cavity detection scheme. In particular, Bogorad \emph{et al.} calculated the number of photons converted to the signal frequency $\omega_s$ by the EH interaction~\cite{bog19}. Scaled in terms of standard parameters the EH interaction generates a low yield of singal photons even for long measurement integration times. We confirmed their result given below,
\begin{equation}\label{eq-N_EH}
N_{\rm EH}=3.1
\left(\frac{Q_s}{10^{12}}\right)^2\left(\frac{E_0}{45\,{\rm MV/m}}\right)^6 
\frac{1-{\rm e}^{-\frac{\omega_3^\infty}{Q_s}t_p}}{\omega_3^\infty t_p/Q_s} 
\times
\left(
\frac{\omega_3^\infty t_p/2Q_s}{1-{\rm e}^{-\frac{\omega_3^\infty}{2Q_s}t_p}}
\frac{1-{\rm e}^{-\frac{\omega_1^\infty}{Q_1}t_p-\frac{\omega_2^\infty}{2Q_2}t_p}}{\omega_1^\infty t_p/Q_1+\omega_2^\infty t_p/2Q_2} 
\right)^2.
\end{equation}

Figure \ref{fig-N} shows the predicted number of photons generated by NLM, leakage, and EH sources at the EH signal frequency $\omega_s$: $N_{\rm NLM}$, $N_{\rm leak}$, and $N_{\rm EH}$, as a function of the pump field strength $E_0$ for cavities with the quality factors, $Q_s=10^{12}$ and $10^{22}$, and pumping periods times, $t_p=5\times10^{15}/\omega_s\approx t_{\rm int}$ and $100/\omega_s$. 

\begin{figure}[h]
\begin{center}
\includegraphics[width=0.9\columnwidth]{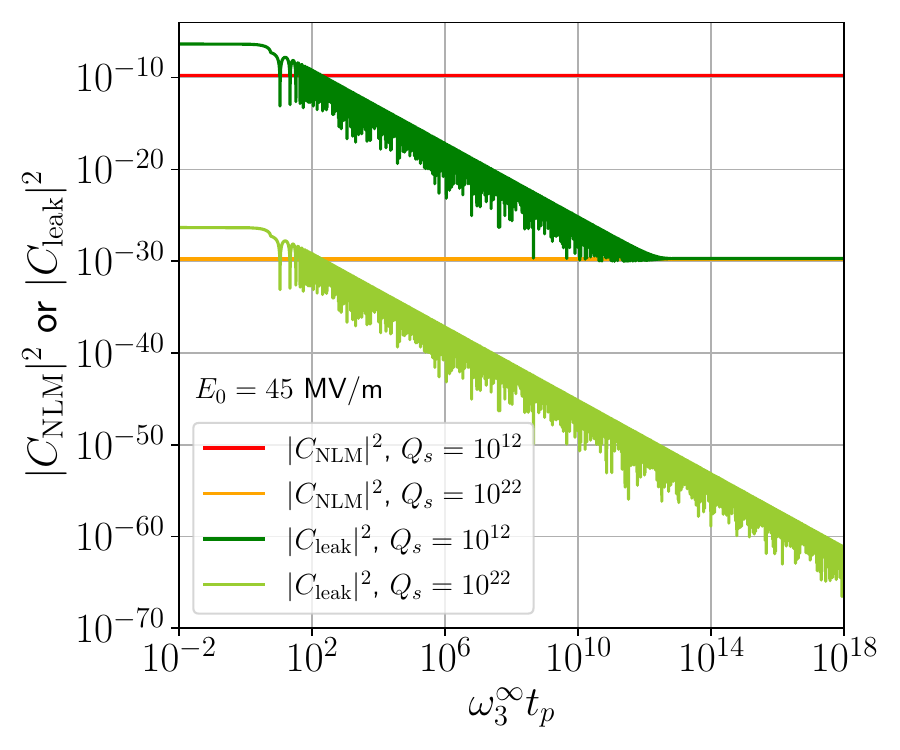}
\caption{Coefficients in Eqs.~\eqref{eq-C_NLM+leak} for the number of NLM photons for cavities with $Q_s=10^{12}$ (red line) and $10^{22}$ (orange line). 
Similarly, for the leakage noise: $Q_s=10^{12}$ (green line) and $10^{22}$ (yellow green line). 
The pump field strength was set to $E_0=45$ MV/m.}
\label{fig-C}
\end{center}
\end{figure}

Note that photon conversion via the NLM effect, $N_{\rm NLM}$, dominates the leakage of pump photons into the signal mode, $N_{\rm leak}$, for the long period case, $t_p\approx t_{\rm int}$. In this limit the spectral function, $|C_{\rm leak}|^2$, is suppressed due to the frequency difference of order GHz between $\omega_{\rm leak}$ and $\omega_2$ (See Eqs.~\eqref{eq-C_leak} and~\eqref{eq-Ci}).
However, for short periods, $t_p=100/\omega_s$, the denominator that suppresses $C_{\rm leak}$ is cancelled, and the spectral weight $|C_{\rm leak}|^2$ becomes larger than $|C_{\rm NLM}|^2$. This cross over is highlighted in Fig.~\ref{fig-C} for the spectral weights, $|C_{\rm leak}|^2$ and $|C_{\rm NLM}|^2$ as a function of $t_p$ for cavities with $Q_s=10^{12}$ and $10^{22}$ at high pump power, $E_0=45$ MV/m. Note that $|C_{\rm leak}|^2$ is dominates $|C_{\rm NLM}|^2$ only for short pump cycles. 
The predicted oscillations in Eq.~\eqref{eq-Ci} for $|C_{\rm leak}|^2$ as a function of $1\ll\omega_3^\infty t_p\ll Q_s$ are evident in Fig.~\ref{fig-C}, as well as the upper right panel of Fig.~\ref{fig-N}.

\begin{figure*}[t]
\begin{center}
\includegraphics[width=0.8\linewidth]{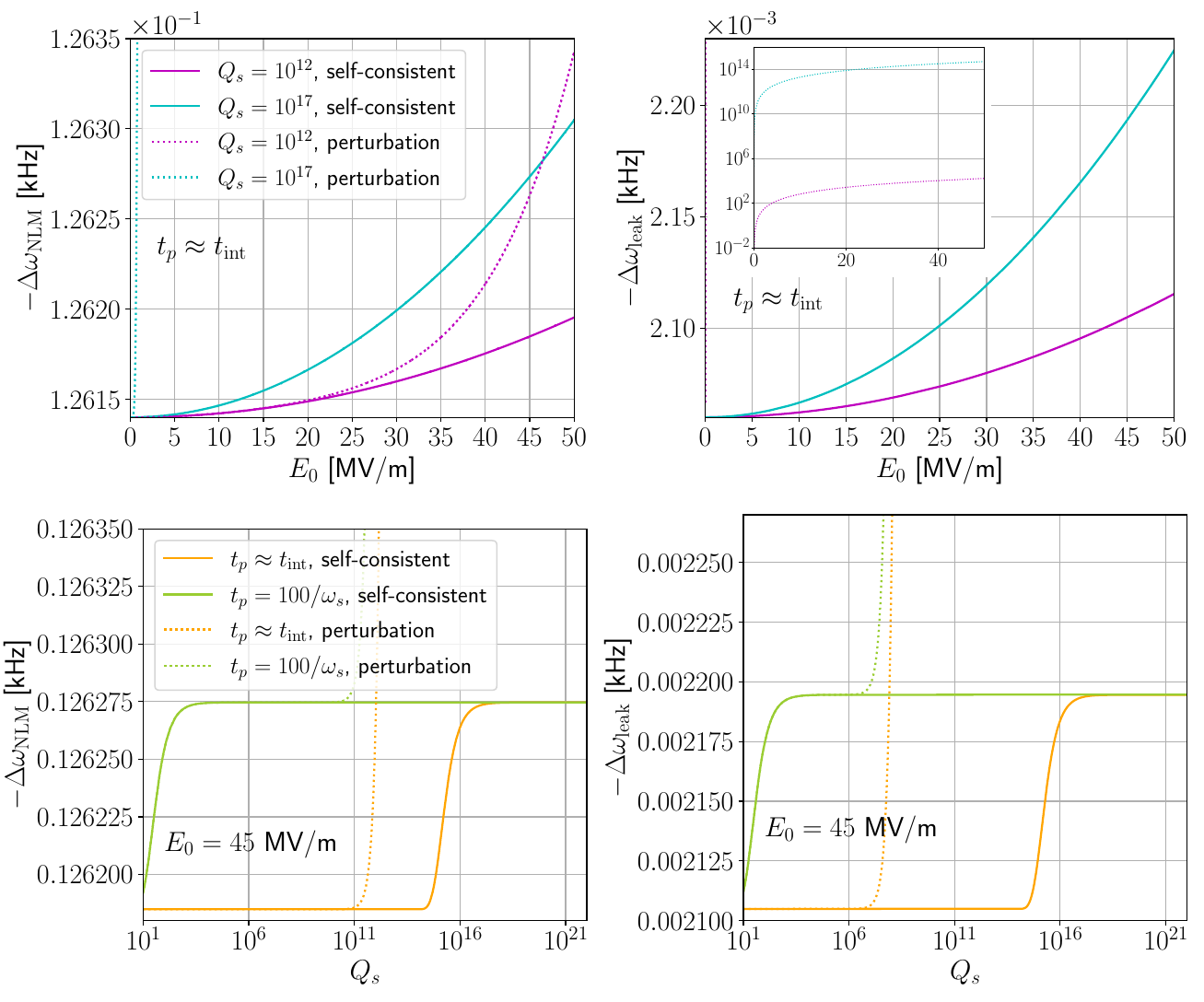}
\caption{Frequencies of photons generated by the NLM current (left panels) and photons leaking from the pump modes into the signal mode (right panels) relative to the axion or EH signal frequency at $\omega_s$:
$\omega_{\rm NLM}\equiv\omega_s+\Delta\omega_{\rm NLM}$ and 
$\omega_{\rm leak}\equiv\omega_s+\Delta\omega_{\rm leak}$.
The upper row shows the dependence of the frequency shifts on the pump field strength $E_0$ for quality factors, $Q_s=10^{12}$ (magenta lines) and $10^{17}$ (cyan lines), and pumping periodic time, $t_p=5\times10^{15}/\omega_s\approx t_{\rm int}$. The inset shows the failure of the perturbation theory result for the frequency shift for high field amplitudes. The lower row shows the dependence of the frequency shifts on $Q_s$ for cavities with pumping periods, $t_p\approx t_{\rm int}$ (orange lines) and $100/\omega_s$ (yellow green lines) at $E_0=45$ MV/m. The solid (dotted) lines are based on a self-consistent (perturbative) calculation described in Sec.~\ref{sec:iid}.
}
\label{fig-dome}
\end{center}
\end{figure*}

For ultra-high Q cavities, $Q_s=10^{22}$, shown in the right panels of Fig.~\ref{fig-N}, the generation of signal photons by the EH interaction exceeds false positive photons from the NLM effect. However, leakage of pump photons into the signal mode still overwhelms photon generation by EH interaction for short pulse times. For long pulse times and ultra-high Q the production of photons by the EH interaction exceeds false positives from both the NLM effect and the leakage noise.
A key to understanding the power dependence of photon conversion from the NLM and leakage fields, and thus the suppression of false positive signals for detection of the EH interaction is the frequency of photons generated by the NLM and leakage fields.
Figure~\ref{fig-dome} shows the frequency \emph{shifts} of the distributions of photons generated by the NLM effect and the leakage from the pump modes into the signal mode, relative to that of photons generated by the EH effect at the signal frequency $\omega_s$,
i.e. 
$\Delta\omega_{\rm NLM}\equiv\omega_{\rm NLM}-\omega_s$
$\Delta\omega_{\rm leak}\equiv\omega_{\rm leak}-\omega_s$,
as functions of $E_0$ and $Q_s$ for $t_p\approx t_{\rm int}$ and $100/\omega_s$. 

\bigskip

\twocolumngrid

\eject

\begin{figure}[h]
\begin{center}
\includegraphics[width=0.9\columnwidth]{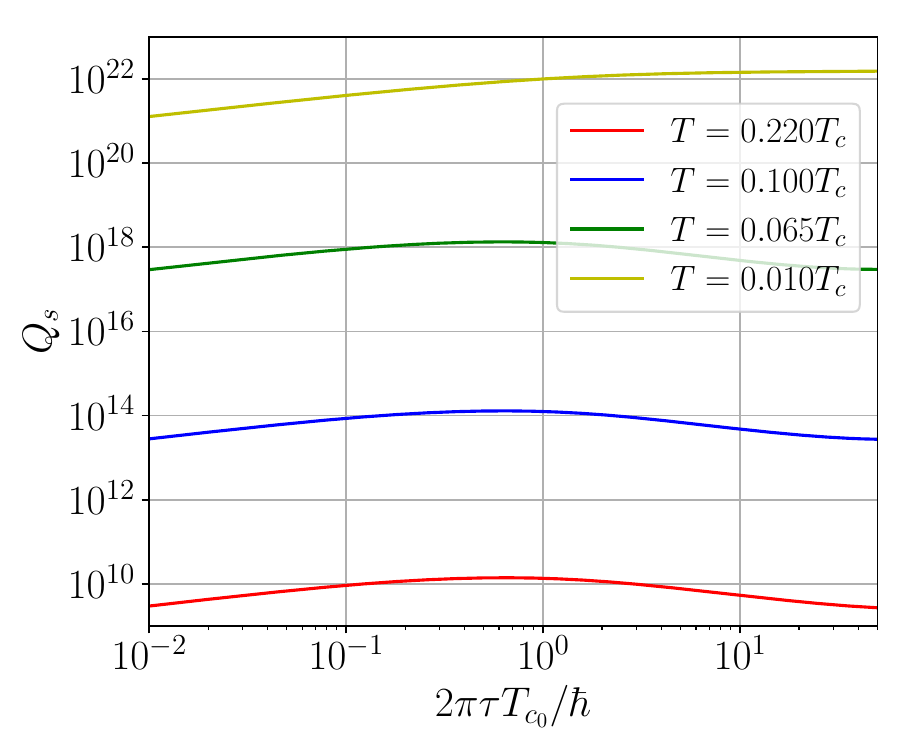}
\caption{Impurity dependence of the quality factor with Nb SRF cavities in the linear response limit at temperatures, $T=0.220T_c$ (red line), $0.100T_c$ (blue line), $0.065 T_c$ (green line), and $0.010T_c$ (yellow line).}
\label{fig-Q}
\end{center}
\end{figure}

\onecolumngrid

We calculate the frequency shifts by solving Eqs.~\eqref{eq-nonlinear_delta_omega_NLM},~\eqref{eq-nonlinear_delta_omega} and~\eqref{eq-form_EM_pump} self-consistently, and for comparison we include the perturbative (non-self-consistent) value for the shifts obtained by replacing the surface magnetic fields on the right-hand side of Eqs.~\eqref{eq-nonlinear_delta_omega_NLM} and~\eqref{eq-nonlinear_delta_omega} with fields that do not include the frequency shifts. 
The frequency shifts obtained from a self-consistent calculation exhibit weaker dependence on the pump field strength, and are nearly independent of $Q_s$ except for the the small change ``jump'' at $Q_s\approx\omega_3^\infty t_p$ [c.f. Eqs.~\eqref{eq-nonlinear_delta_omega_NLM} and~\eqref{eq-nonlinear_delta_omega}
]. 
By contrast the perturbative result for the frequency shifts diverge at high $Q_s$ and $E_0$ signalling the failure of the perturbation calculation for high pump power and high $Q_s$.

\begin{figure*}[t]
\begin{center}
\includegraphics[width=0.8\linewidth]{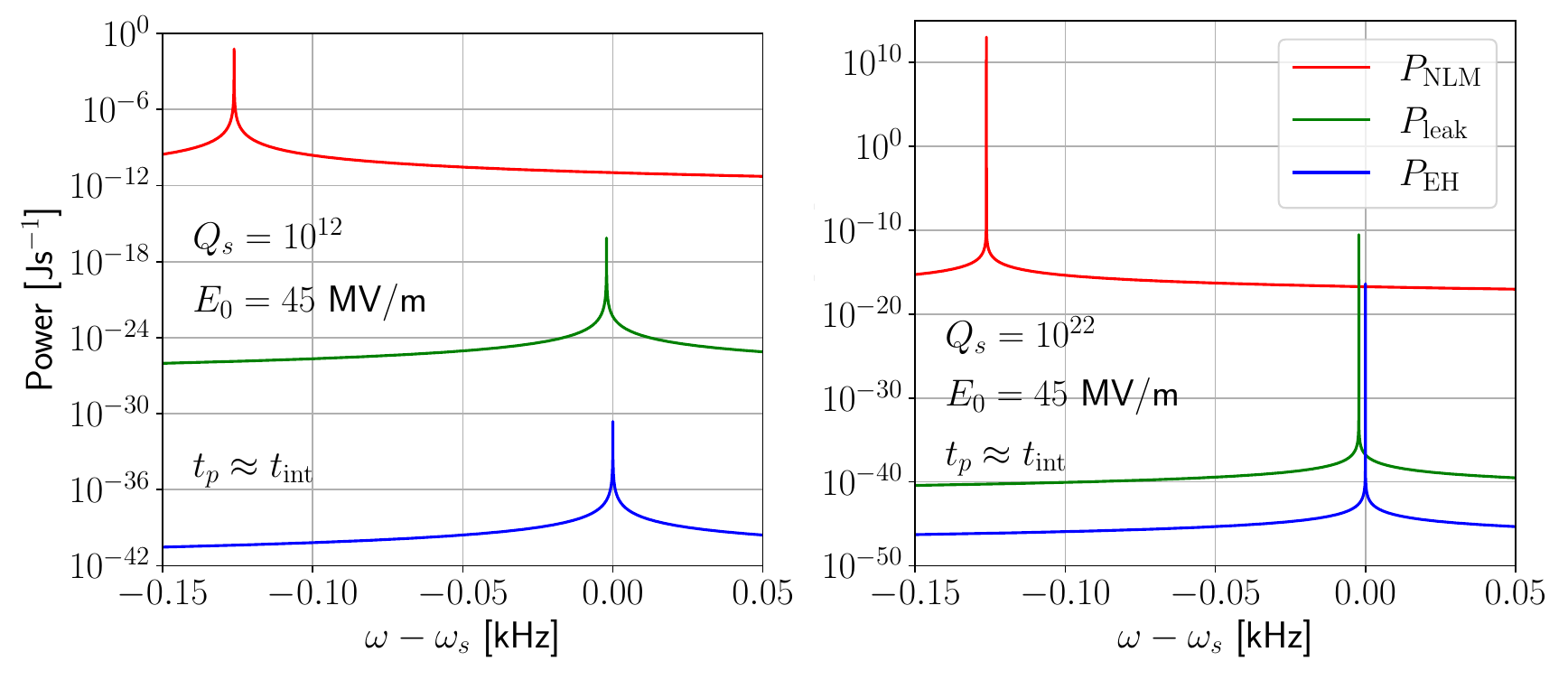}
\caption{Power spectra for NLM radiation (red lines), the EH interaction (blue lines) and leakage radiation (green lines) as a function of frequency relative to the EH signal frequency $\omega_s$ for cavities with $Q_s=10^{12}$ (left panel) and $10^{22}$ (right panel) at $E_0=45$ MV/m. The peaks are the exact maximum peak values of each power spectrum.}
\label{fig-P} 
\end{center}
\end{figure*}

We also calculated the power spectra as a function of frequency 
$P_{\rm NLM}(\omega)=(\hbar\omega^2/Q_s)\,N_{\rm NLM}$,
$P_{\rm leak}(\omega)=(\hbar\omega^2/Q_s)\,N_{\rm leak}$,
$P_{\rm EH}(\omega)=(\hbar\omega^2/Q_s)\,N_{\rm EH}$,
where the expressions for $N_{\rm NLM}$, $N_{\rm leak}$, and $N_{\rm EH}$ are evaluated as 
functions of $\omega$~\cite{padamsee08,gao21}. 
Figure \ref{fig-P} shows the power spectra for photons generated by the NLM, leakage and EH sources 
for SRF cavities with $Q_s=10^{12}$ and $10^{22}$, $E_0=45$ MV/m, and $t_p\approx t_{\rm int}$.
The maximum (``peak'') positions of the power spectra from these different sources separate due to the different frequency shifts. As $Q_s$ increases the frequency shift remain constant, but the contribution of the EH term begins to dominate at $\omega=\omega_s$ as the spectra become 
sharper.

\subsection{Sensitivity to Axion Generated Signal Photons\label{sec:iiib}} 

The generation of photons by the NLM effect and the leakage of photons from the pump modes into the signal mode also impact the sensitivity of the cavity to detection of possible axion-mediated photons, and thus place a limit on the sensitivity to the axion-photon interaction strength, $g_{a\gamma\gamma}$. 
Berlin {\it et al.} considered several background noise sources including thermal photons at the signal frequency. The thermal background is given by $N_{\rm th}\equiv k_{\rm B}T/\hbar\omega_s$, where $k_{\rm B}$ is the Boltzmann constant~\cite{ber20}. 
Here we analyze the limiting sensitivity to the axion-photon coupling including the  NLM background, leakage noise and thermal noise. 
We replace $N_{\rm th}$ in Eq. (17) of Ref.~\cite{bog19} with $N_{\rm th}+N_{\rm NLM}+N_{\rm leak}$, and consider the dependence of the limiting axion-photon coupling as functions of pump mode amplitude, $E_0$,and quality factor, $Q_s$. 
Other parameters such as the pump period and signal frequency are the same as those used in the calculation of the number of photons generated by the NLM effect, leakage noise and EH interaction to generate Fig.~\ref{fig-N}. Then, the lower limit of sensitivity to the light axion-photon interaction strength, defined as $g_{a\gamma\gamma}$ for light axions at signal-to-noise ratio, SNR=1, becomes~\cite{bog19,kah22},
\begin{align} 
&g_{a\gamma\gamma,0}^{\rm lim} 
=1.6\times10^{-11} \ {\rm GeV}^{-1}
\times\left(\frac{10^{12}}{Q_s}\right)^{1/4} 
\left(\frac{45 \ {\rm MV/m}}{E_0}\right)^{3/2}
\left(20 \ {\rm days}\times B\right)^{1/8}
\times
\left(\frac{N_{\rm th}+N_{\rm NLM}+N_{\rm leak}}{N_{\rm th}}\right)^{1/4}
\notag \\
&
\qquad\qquad\qquad
\qquad\qquad\qquad
\times
\left(\frac{\omega_3^\infty t_p/Q_s}{1-{\rm e}^{-\frac{\omega_3^\infty}{Q_s}t_p}}\right)^{1/4} 
\times
\left(
\frac{1-{\rm e}^{-\frac{\omega_3^\infty}{2Q_s}t_p}}{\omega_3^\infty t_p/2Q_s}
\frac{\omega_1^\infty t_p/Q_1+\omega_2^\infty t_p/2Q_2}{1-{\rm e}^{-\frac{\omega_1^\infty}{Q_1}t_p-\frac{\omega_2^\infty}{2Q_2}t_p}} 
\right)^{1/2}, 
\end{align}
where the cavity bandwith $B$ is given by
\begin{align} 
B=
\left\{ \begin{array}{ll} 
\displaystyle \omega_3^\infty/Q_s,  & \omega_3^\infty t_p\gg Q_s, \\
\displaystyle 1/t_{\rm int}, & \omega_3^\infty t_p\ll Q_s.
\end{array}\right. 
\end{align}

\begin{figure*}[t]
\begin{center}
\includegraphics[width=0.8\linewidth]{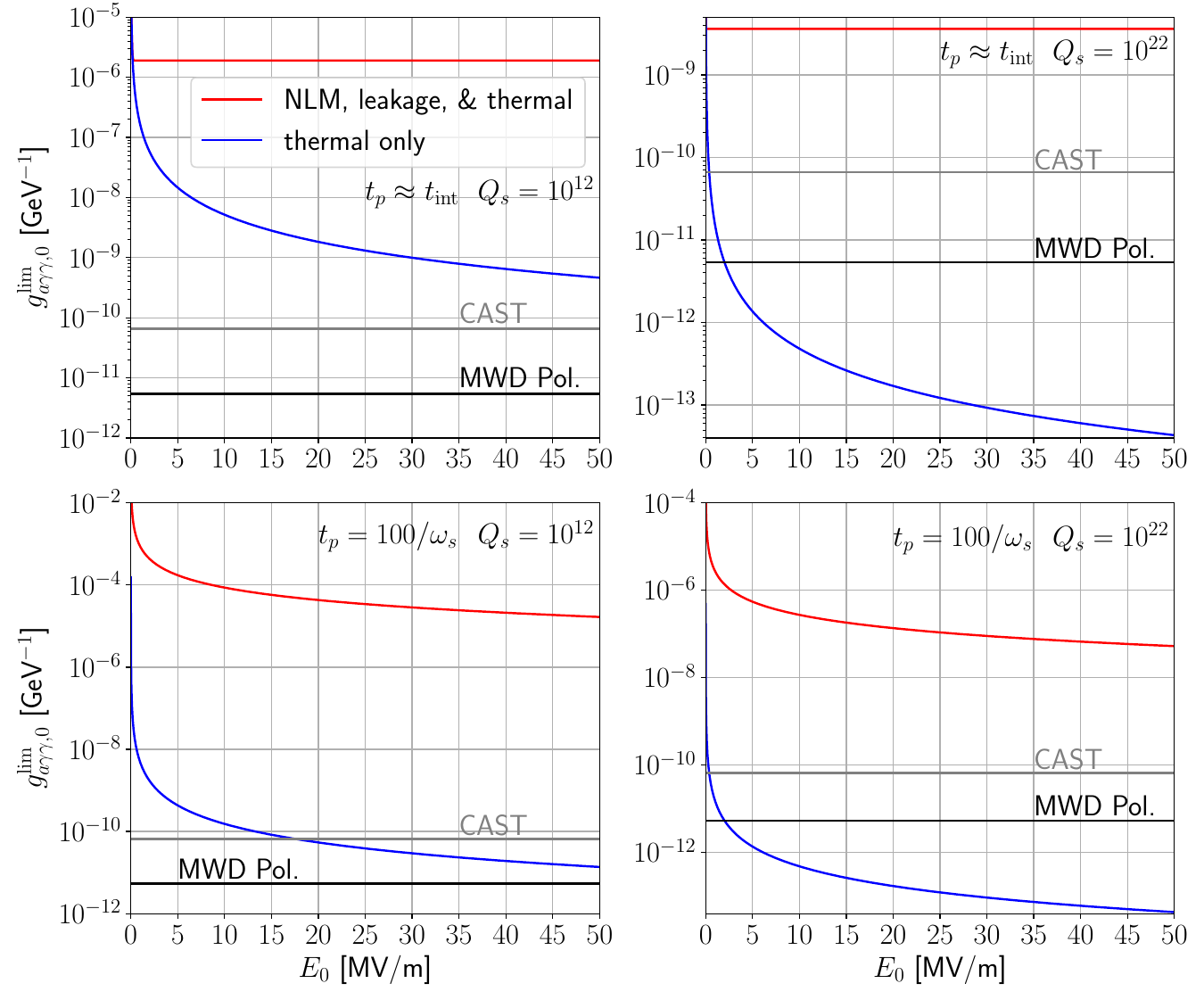}
\caption{Sensitivity to the light axion-photon interaction strength $g_{a\gamma\gamma,0}^{\rm lim}$ with combined background photons from NLM radiation, leakage noise and thermal noise (red lines) compared with only thermal noise (blue lines) as a function of pump field strength $E_0$ for the quality factors $Q_s=10^{12}$ at $T=1.5$ K (left panels) and $Q_s=10^{22}$ at $T=0.1$ K (right panels), and pumping periods, $t_p=5\times10^{15}/\omega_s\approx t_{\rm int}$(upper row) and $100/\omega_s$ (lower row). The grey and black lines are the limits $g_{a\gamma\gamma}<6.6\times10^{-11}$ GeV$^{-1}$ and $5.4\times10^{-12}$ GeV$^{-1}$ obtained from solar axion searches with the CERN Axion Solar Telescope~\cite{ana17} and polarization measurements of thermal radiation from magnetic white dwarf stars~\cite{des22}, respectively.}
\label{fig-g}
\end{center}
\end{figure*}

Results for the limiting value of $g_{a\gamma\gamma,0}^{\rm lim}$ as a function of the pump field amplitude based on the thermal background are compared with the result based on the combined backgrounds from the NLM effect, leakage noise and thermal noise in Fig.~\ref{fig-g}.
The limits are shown as functions of the pump field amplitude $E_0$ for the quality factors $Q_s=10^{12}$ at $T=1.5$ K and $Q_s=10^{22}$ at $T=0.1$ K, and for pump periods of $t_p\approx t_{\rm int}$ and $100/\omega_s$. 
Also shown are the limits on $g_{a\gamma\gamma}<6.6\times10^{-11}$ GeV$^{-1}$ and $5.4\times10^{-12}$ GeV$^{-1}$ that were obtained from solar axion searches with the CERN Axion Solar Telescope (CAST)~\cite{ana17} and magnetic white dwarf polarization (MWD Pol.) measurements~\cite{des22}, respectively. 

The NLM background, proportional to $E_0^6$, is dominant for $t_p\approx t_{\rm int}$, in which case $g_{a\gamma\gamma,0}^{\rm lim}$ is nearly constant since the number of photons generated by a light axion is also proportional to $E_0^6$. For short periods, $t_p=100/\omega_s$, the leakage noise dominates, and $g_{a\gamma\gamma,0}^{\rm lim}$ varies with $E_0$ as $N_{\rm leak}$, i.e. a polynomial with $E_0^2$, $E_0^4$ in addition to $E_0^6$.
A setup with $\omega_3^\infty t_p\gg Q_s$, i.e. $B=\omega_3^\infty/Q_s$, removes the leakage noise, but also weakens photon conversion by the axion field and EH interaction~\cite{bog19}.
In terms of the potential for detecting the axion field the NLM effect is most problematic; the NLM background cannot be removed in the single cavity setup, and thus the limit for the axion-photon coupling is significantly worse than previously estimated. The bottom line is that the NLM background and leakage noise significantly impact the sensitivity to light axions, with $g_{a\gamma\gamma,0}^{\rm lim}$ above the CAST and MWD Polarization limits over the full range of parameters we explored.

\subsection{Detecting the Axion Field with Two SRF Cavities}

Lastly we discuss the two-cavity setup proposed by Gao and Harnik~\cite{gao21}. We show that it may be suitable for searching for photon conversion by the axion field. Their method is summarized as follows: 
\begin{itemize}\setlength\itemsep{-0.1em}
\item Two identical SRF cavities, an ``emitter'' and a ``receiver'' cavity are required.
\item Two modes, e.g. a TM mode (mode $0$) and a TE mode (mode $1$) with ${\bf E}\cdot{\bf B}\ne 0$ and mode frequencies $\omega_0$ and $\omega_1$, are pumped with photons.
\item A receiver cavity - identical to the emitter cavity - is pumped in mode $0$ while mode $1$ 
is otherwise unpopulated, and thus serves as the signal detection mode.
\item The axion field is source by the two pump modes in the emitter cavity at frequencies $\omega_\pm=|\omega_0\pm\omega_1|$.
\item An axion at $\omega_1-\omega_0$ can covert to a photon in the receiver cavity with frequency $\omega_1$ providing detection of photon conversion by the axion field created in the emitter.
\end{itemize}

This method avoids the generation of a large population of background photons at the signal frequency from the NLM effect since the screening current of the pump mode in the receiver cavity generates NLM photons only at $3\omega_0$, which is neither a resonant mode of the receiver, and can be chosen to be far from resonance with the signal mode at $\omega_1$.
The two-cavity method can also avoid leakage noise of the spectator mode at $\omega_0$ into the signal mode at $\omega_1$. In particular, the leakage noise field is proportional to the surface integral 
\begin{align} 
{\bf E}_{\rm leak}({\bf r})\propto{\bf e}_1({\bf r})
\int_S[{\bf E}_0({\bf r}')\times{\bf b}_1({\bf r}')]\cdot d{\bf a}'
\,, 
\end{align}
We obtain $({\bf E}_0\times{\bf b}_1)\cdot d{\bf a}=0$ if we choose mode $0$ as ${\rm TE}_{011}$ and mode $1$ as ${\rm TM}_{010}$ or vice versa, as Gao and Harnik propose. As a result we can use the two-cavity setup with $B=1/t_{\rm int}$ and zero leakage noise. 

The heterodyne detection of axion DM with an SRF cavity as proposed by Berlin \emph{et al.}~\cite{ber20,ber21} is similar in terms of noise suppression to the two-cavity setup with the emitter cavity ``replaced'' by the universe.
Thus, Berlin's method can also avoid photon conversion by the NLM effect, $B=1/t_{\rm int}$ and no leakage noise. 
It is important to note that we cannot use the setup with $B=1/t_{\rm int}$ without the leakage noise for the TE mode into higher frequency TE modes discussed in App. A of Ref.~\cite{ber20}.
Lastly, we note that the EH interaction allows for photon conversion from the pumped spectator mode at $\omega_0$ to the third harmonic at $3\omega_0$. However, detection of the photon conversion by the EH interaction for current state of the art cavities is not feasible because the NLM effect will 
generate photons at $3\omega_0$.

\section{Summary and Challenges\label{sec:iv}} 

\underline{Summary:} 
High-Q superconducting resonators provide promising platforms for search and detection of physics beyond the standard model, including DM candidates, light-by-light scattering in QED, and gravitational waves. Essential to many of these detection schemes is photon frequency conversion.
Superconducting resonators are nonlinear devices. As a result the screening currents that confine photons in the cavity also generate frequency conversion of microwave photons in the cavity.
We consider photon frequency conversion for nonlinear axion-electrodynamics, the EH interaction, and Meissner radiation in high-Q superconductoing (SRF) cavities.   
In a single SRF cavity with two pump modes photon frequency conversion by the Meissner screening current dominates photon generation by the EH interaction for cavities with $Q \lesssim 10^{12}$. 
Meissner currents also generate background photons that limit the operation of the resonator for axion detection in three-mode, single cavity setups. 
Leakage of photons from pump modes into the signal mode for detection of both axion and EH mediated light-by-light scattering is also considered. 
We find that photon frequency conversion by the EH interaction can exceed Meissner and leakage radiation for \emph{ultra-high-Q} cavities at low temperatures; requirements that are beyond current state of the art cavities. 

\underline{Challenges:} 
The theory of nonequilibrium superconductivity for weakly disordered fully gapped superconductors provides predictions and targets for possible ultra-high Q superconducting cavity sensors.
In the case of the multi-mode, single-cavity setup surpressing the background of photons generated by the NLM current and leakage noise is a challenge because the materials and cryogenic requirements are beyond current state of the art.
In particular Fig.~\ref{fig-Q} shows the quality factor for Nb SRF cavities in the linear response limit as a function of the quasiparticle scattering time $\tau$ calculated using the method described in Ref.~\cite{uek22}, where $T_c$ is the transition temperature of the disordered superconductor. Note that to reach $Q_s\sim10^{22}$ requires the removal of TLS losses from niobium oxides as well as subgap states from disorder induced pair-breaking in the superconducting material, as well as operating the SRF cavity at very low temperatures, $T\approx0.01T_c\approx0.1$ K for Nb. Both the material purity and the cryogenic requirement in the presence of high power microwave pump fields pose major challenges.
The signal power peak is very narrow, $\Delta\omega_s=\omega_s/Q_s$. To capture this peak, shown in Fig.~\ref{fig-P}, the cavity design requires errors of $\Delta L/L$ and $\Delta R/R \lesssim 1/Q_s$. Thus, advances in cavity engineering may be required to address this challenge.

Another challenge arises from the potential frequency shifts due to pressure of the electromagnetic field, referred to as \emph{Lorentz force detuning} in Refs.~\cite{padamsee08} and \cite{aun00}. Table. II of Ref.~\cite{aun00} reports frequency shifts of TTF cavities due to the Lorentz force detuning of $\sim 600\,$Hz. Lorentz force detuning depends on cavity construction, thus, it may be necessary to design cavities in order to suppress the effects of electromagnetic pressure for QED measurements and axion searches, perhaps by engineered reinforcement of cavities.
Lastly, the dual cavity setup for detection of axion mediated photon frequency conversion can avoid significant backgrounds from NLM radiation and leakage noise, but the setup requires two cavities with identical mode spectra, a design and engineering challenge.

\begin{acknowledgments}

We thank Yonatan Kahn and Roni Harnik for discussions and their insights on detecting light by light scattering using high-Q SRF cavities, and Bianca Giaccone for discussions on possible experiments to measure the NLM generated radiation using SRF cavities.
This research was supported by the U.S. Department of Energy, Office of Science, National Quantum Information Science Research Centers, Superconducting Quantum Materials and Systems Center (SQMS) under contract number DE-AC02-07CH11359.
A draft of this manuscript was written during a recent stay at the Aspen Center for Physics which is supported by National Science Foundation grant PHY-2210452.

\end{acknowledgments}

\appendix 

\section{Cavity basis functions for TE and TM modes\label{app:a}} 

We include the TE and TM modes functions for an ideal cylindrical cavity~\cite{gao21,hill09,jackson67}. Assuming that the charge and current densities, $\rho$ and ${\bf j}$, are zero in a hollow cavity, the wave equations for the EM fields in the vacuum region of cavity are given by
\begin{subequations} 
\begin{align} 
\left({\bm\nabla}^2-\frac{1}{c^2}\frac{\partial^2}{\partial t^2} \right){\bf E}({\bf r},t)={\bf 0}, \label{eq-wave_equations_E_ideal} 
\end{align} 
\begin{align} 
\left({\bm\nabla}^2-\frac{1}{c^2}\frac{\partial^2}{\partial t^2} \right){\bf B}({\bf r},t)={\bf 0}. \label{eq-wave_equations_M_ideal}
\end{align}\label{eq-wave_equations_ideal}%
\end{subequations} 

First consider the TE modes with $E_z=0$, and assume the form of $B_z$ is given by 
\begin{align} 
B_z({\bf r},t)=B_z(\rho){\rm e}^{in\varphi}{\rm e}^{ik_zz}{\rm e}^{-i\omega t}. \label{eq-form_Bz}
\end{align}
Substituting Eq. (\ref{eq-form_Bz}) into Eq. (\ref{eq-wave_equations_M_ideal}), we obtain Bessel's equation for $B_z$,
\begin{align} 
\rho^2\frac{d^2B_z(\rho)}{d\rho^2}
+\rho\frac{dB_z(\rho)}{d\rho} 
+
\left\{ 
\left[\left(\frac{\omega}{c}\right)^2
-\left(\frac{l\pi}{L}\right)^2\right]\rho^2
-n^2
\right\}
B_z(\rho)=0, \label{eq-Bessel_Bz}
\end{align}
where $l$ is given by $l= 0,1, 2, \cdots$ so that $B_z$ vanishes at $z=0$ and $L$. 
The solution for $B_z$ to be finite at $\rho=0$ is given by 
\begin{align} 
B_z(\rho)=CJ_n
\left(\sqrt{\left(\frac{\omega}{c}\right)^2
-\left(\frac{l\pi}{L}\right)^2}\rho\right), 
\end{align}
where $C$ is an integration constant. 
Using the boundary condition so that $B_\rho$ and $E_\varphi$ vanish at $\rho=0$, $(\partial B_z/\partial \rho)|_{\rho=R}$, 
we obtain $\omega$ as 
\begin{align} 
\omega=\omega_{nml}^{\rm TE}
=c\sqrt{\left(\frac{S_{nm}}{R}\right)^2
+\left(\frac{l\pi}{L}\right)^2}, 
\end{align}
Here we restrict ourselves to the modes at $n=0$, and then obtain the rest of ${\bf E}$ and ${\bf B}$ from Eqs. (\ref{eq-Faraday-2}) and (\ref{eq-no_magnetic_monopole-2}) and $C=B_z(0)$ for $n=0$ as
\begin{subequations} 
\begin{align} 
{\bf E}_{0ml}^{\rm TE}({\bf r}, t)
=&iB_z(0)
\frac{\omega_{0ml}^{\rm TE}}{c}\frac{R}{S_{0m}}J_1\left(\frac{S_{0m}}{R}\rho\right)\hat{\bm\varphi}\,{\rm e}^{i\frac{l\pi}{L}z-i\omega_{0ml}^{\rm TE}t},
\\
{\bf B}_{0ml}^{\rm TE}({\bf r},t)
=&B_z(0)
\bigg[
-i\frac{l\pi}{L}\frac{R}{S_{0m}}J_1\left(\frac{S_{0m}}{R}\rho\right)\hat{\bm\rho} 
+J_0\left(\frac{S_{0m}}{R}\rho\right)\hat{\bf z}
\bigg]
{\rm e}^{i\frac{l\pi}{L}z-i\omega_{0ml}^{\rm TE}t}. 
\end{align}\label{eq-TE_EM}
\end{subequations} 

We next consider the TM modes at $B_z=0$, and assume that the form of $E_z$ is given by 
\begin{align} 
E_z({\bf r},t)=E_z(\rho){\rm e}^{in\varphi}{\rm e}^{ik_zz}{\rm e}^{-i\omega t}. \label{eq-form_Ez}
\end{align}
Substituting Eq. (\ref{eq-form_Ez}) into Eq. (\ref{eq-wave_equations_E_ideal}), we obtain equation for $E_z$ as 
\begin{align} 
\rho^2\frac{d^2E_z(\rho)}{d\rho^2}
+\rho\frac{dE_z(\rho)}{d\rho}
+
\left\{ 
\left[\left(\frac{\omega}{c}\right)^2
-\left(\frac{l\pi}{L}\right)^2\right]\rho^2
-n^2
\right\}
E_z(\rho)=0, \label{eq-Bessel_Ez}
\end{align}
where $l$ is also given by $l=0, 1, 2, \cdots$ so that $E_z$ vanishes at $z=0$ and $L$.  By solving Eq. (\ref{eq-Bessel_Ez}), $E_z$ to be finite at $\rho=0$ is given by
\begin{align} 
E_z(\rho)=CJ_n
\left(\sqrt{\left(\frac{\omega}{c}\right)^2
-\left(\frac{l\pi}{L}\right)^2}\rho\right). 
\end{align}
We consider boundary condition so that $E_z$ vanishes at $\rho=R$, and then obtain $\omega$ as 
\begin{align} 
\omega=\omega_{nml}^{\rm TM}
=c\sqrt{\left(\frac{Z_{nm}}{R}\right)^2
+\left(\frac{l\pi}{L}\right)^2}, 
\end{align}
Again, we restrict ourselves to the modes at $n=0$, and then obtain the rest of ${\bf E}$ and ${\bf B}$ from Eqs. (\ref{eq-Faraday-2}) and (\ref{eq-Gauss-2}) and $C=E_z(0)$ for $n=0$ as 
\begin{subequations} 
\begin{align} 
{\bf E}_{0ml}^{\rm TM}({\bf r}, t)  
=&E_z(0)
\bigg[
-i\frac{l\pi}{L}\frac{R}{Z_{0m}}J_1\left(\frac{Z_{0m}}{R}\rho\right)\hat{\bm\rho} 
+J_0\left(\frac{Z_{0m}}{R}\rho\right)\hat{\bf z}
\bigg]
\times {\rm e}^{i\frac{l\pi}{L}z-i\omega_{0ml}^{\rm TM}t}, 
\\
{\bf B}_{0ml}^{\rm TM}({\bf r}, t)
=&-iE_z(0)
\omega_{0ml}^{\rm TM}\frac{R}{Z_{0m}}J_1\left(\frac{Z_{0m}}{R}\rho\right)\hat{\bm\varphi} 
\times{\rm e}^{i\frac{l\pi}{L}z-i\omega_{0ml}^{\rm TM}t}. 
\end{align}\label{eq-TM_EM}
\end{subequations} 

The TE$_{011}$ and TM$_{010}$ EM fields can be chosen from Eqs. (\ref{eq-TE_EM}) and (\ref{eq-TM_EM}) as \cite{bog19}
\begin{subequations} 
\begin{align} 
{\bf E}_{011}^{\rm TE}({\bf r}, t)
=&iB_z(0)
\frac{\omega_{011}^{\rm TE}}{c}\frac{R}{S_{01}}J_1\left(\frac{S_{01}}{R}\rho\right)\sin\frac{\pi z}{L}\hat{\bm\varphi} 
\times\sin\omega_{011}^{\rm TE}t\,,
\\
{\bf B}_{011}^{\rm TE}({\bf r}, t)  
=&iB_z(0)
\bigg[
-\frac{R}{S_{01}}J_1\left(\frac{S_{01}}{R}\rho\right)\frac{\pi}{L}\cos\frac{\pi z}{L}\hat{\bm\rho} 
+J_0\left(\frac{S_{01}}{R}\rho\right)\sin\frac{\pi z}{L}\hat{\bf z}
\bigg]
\cos\omega_{011}^{\rm TE}t, 
\\
{\bf E}_{010}^{\rm TM}({\bf r}, t)  
=&E_z(0)
J_0\left(\frac{Z_{01}}{R}\rho\right)\hat{\bf z}
\cos\omega_{010}^{\rm TM}t, 
\\
{\bf B}_{010}^{\rm TM}({\bf r}, t)
=&-E_z(0)
J_1\left(\frac{Z_{01}}{R}\rho\right)\hat{\bm\varphi} 
\sin\omega_{010}^{\rm TM}t. 
\end{align}\label{eq-TE011_TM010}%
\end{subequations} 
Using the average of the electric field given by $E_0\equiv\sqrt{(1/V)\int dv|{\bf E}_{011}^{\rm TE}({\bf r})|^2}=\sqrt{(1/V)\int dv|{\bf E}_{010}^{\rm TM}({\bf r})|^2}$, we can express Eq. (\ref{eq-TE011_TM010}) as 
\begin{subequations} 
\begin{align} 
&{\bf E}_{011}^{\rm TE}({\bf r}, t)
=\frac{\sqrt{2}E_0}{J_0(S_{01})}
J_1\left(\frac{S_{01}}{R}\rho\right)\sin\frac{\pi z}{L}\hat{\bm\varphi} 
\sin\omega_{011}^{\rm TE}t, 
\\
&{\bf B}_{011}^{\rm TE}({\bf r}, t)  
=\frac{\sqrt{2}E_0}{J_0(S_{01})}
\bigg[
-\frac{c}{\omega_{011}^{\rm TE}}J_1\left(\frac{S_{01}}{R}\rho\right)\frac{\pi}{L}\cos\frac{\pi z}{L}\hat{\bm\rho} 
+\frac{c}{\omega_{011}^{\rm TE}}\frac{S_{01}}{R}J_0\left(\frac{S_{01}}{R}\rho\right)\sin\frac{\pi z}{L}\hat{\bf z}
\bigg]
\cos\omega_{011}^{\rm TE}t, 
\\
&{\bf E}_{010}^{\rm TM}({\bf r}, t)  
=\frac{E_0}{J_1(Z_{01})}
J_0\left(\frac{Z_{01}}{R}\rho\right)\hat{\bf z}
\cos\omega_{010}^{\rm TM}t, \\
&{\bf B}_{010}^{\rm TM}({\bf r}, t)
=-\frac{E_0}{J_1(Z_{01})}
J_1\left(\frac{Z_{01}}{R}\rho\right)\hat{\bm\varphi} 
\sin\omega_{010}^{\rm TM}t. 
\end{align}\label{eq-TE011_TM010-2}%
\end{subequations} 
We adopt Eq. (\ref{eq-TE011_TM010-2}) as the pump fields. 

\section{Maxwell's equations in a hollow cavity\label{app:b}} 

We consider the EM fields in a hollow cavity, and calculate the EM fields based on Slater's method \cite{sla46}.  Starting with the general principles of vector analysis, we can expand the EM fields, and charge and current density as (See also Ch. 3 in Ref. \cite{sla46})
\begin{subequations} 
\begin{align} 
{\bf E}({\bf r},t)=\sum_i&\bigg[{\bf e}_i
({\bf r})\int{\bf E}({\bf r}',t)\cdot{\bf e}_i
({\bf r}')dv'
+{\bf f}_i
({\bf r})\int{\bf E}({\bf r}',t)\cdot{\bf f}_i
({\bf r}')dv'\bigg], \label{eq-expansion_EM1E} 
\\
{\bf B}({\bf r},t)=\sum_i&{\bf b}_i
({\bf r})\int{\bf B}({\bf r}',t)\cdot{\bf b}_i
({\bf r}')dv', 
\\
\rho({\bf r},t)=\sum_i&\psi_i({\bf r})\int\rho({\bf r}',t)\psi_i({\bf r}')dv', 
\\
{\bf j}({\bf r},t)=\sum_i&\bigg[{\bf e}_i
({\bf r})\int{\bf j}({\bf r}',t)\cdot{\bf e}_i
({\bf r}')dv' 
+{\bf f}_i
({\bf r})\int{\bf j}({\bf r}',t)\cdot{\bf f}_i
({\bf r}')dv'\bigg],
\end{align}\label{eq-expansion_EM1}%
\end{subequations} 
where ${\bf e}_i({\bf r})$ and ${\bf b}_i({\bf r})$ are the solenoidal vector functions, which satisfy $k_i{\bf e}_i ={\bm\nabla}\times{\bf b}_i $ and $k_i{\bf b}_i ={\bm\nabla}\times{\bf e}_i $, and ${\bf f}_i({\bf r})$ is the irrotational vector function, which is the gradient of a scalar function $\psi_i({\bf r})$ as $k_i{\bf f}_i ={\bm\nabla}\psi_i$, and $k_i$ is the propagation constant associated with the $i$th mode in an ideal cavity. These functions also satisfy the normalization and orthogonality conditions as 
\begin{subequations} 
\begin{align} 
\int{\bf e}_i
\cdot{\bf e}_j
dv&=\delta_{ij}, \ \ \ 
\int{\bf b}_i
\cdot{\bf b}_j
dv=\delta_{ij}, \label{eq-normalization_orthogonality_sole}\\
\int{\bf f}_i
\cdot{\bf f}_j
dv&=\delta_{ij}, \ \ \ 
\int\psi_i\psi_jdv=\delta_{ij}, \\
\int{\bf e}_i
\cdot{\bf f}_j
dv&=0, 
\end{align}
\end{subequations} 
the wave equations as 
\begin{subequations} 
\begin{align} 
{\bm\nabla}^2{\bf e}_i
+k_i^2&{\bf e}_i
={\bf 0}, \ \ \
{\bm\nabla}^2{\bf b}_i
+k_i^2{\bf b}_i
={\bf 0}, \\
{\bm\nabla}^2{\bf f}_i
+k_i^2&{\bf f}_i
={\bf 0}, \ \ \
{\bm\nabla}^2\psi_i+k_i^2\psi_i=0, 
\end{align}\label{eq-wave_ideal}%
\end{subequations} 
and the boundary conditions as 
\begin{subequations} 
\begin{align} 
&{\bf n}\times{\bf e}_i
={\bf 0}, \ \ \ 
{\bf n}\cdot{\bf b}_i
=0, \ \ \ 
{\bf n}\times{\bf f}_i
={\bf 0}, \ \ \ 
\psi_i=0 \ \ \ {\rm on \ S}, \\
&{\bf n}\times{\bf b}_i
={\bf 0}, \ \ \ 
{\bf n}\cdot{\bf e}_i
=0, \ \ \ 
{\bf n}\times{\bf f}_i
={\bf 0}, \ \ \ 
\psi_i=0 \ \ \ {\rm on \ S'},
\end{align}\label{eq-boundary_ideal}%
\end{subequations} 
where $\bf n$ is the outer normal to the surface, and $S$ and $S'$ are the conductive and insulating surfaces, respectively.  

We can also expand ${\bm\nabla}\times{\bf E}$, ${\bm\nabla}\times{\bf B}$, and ${\bm\nabla}\cdot{\bf E}$ in terms of the functions ${\bf e}_i $, ${\bf b}_i $, ${\bf f}_i $, and $\psi_i$ as 
\begin{subequations} 
\begin{align} 
&{\bm\nabla}\times{\bf E}
=\sum_i{\bf b}_i
\int({\bm\nabla}\times{\bf E})\cdot{\bf b}_i
dv 
=\sum_i{\bf b}_i
\left[k_i\int{\bf E}\cdot{\bf e}_i
dv
+\int_S({\bf E}\times{\bf b}_i)\cdot
d{\bf a}\right], 
\\
&{\bm\nabla}\times{\bf B}
=\sum_i{\bf e}_i
\int({\bm\nabla}\times{\bf B})\cdot{\bf e}_i
dv 
=\sum_i{\bf e}_i
\left[k_i\int{\bf B}\cdot{\bf b}_i
dv
+\int_{S'}({\bf B}\times{\bf e}_i)\cdot
d{\bf a}\right], 
\\
&{\bm\nabla}\cdot{\bf E}
=\sum_i\psi_i\int({\bm\nabla}\cdot{\bf E})\psi_idv 
=-\sum_i\psi_ik_i\int{\bf E}\cdot{\bf f}_i
dv,
\end{align}\label{eq-expansion_EM2}%
\end{subequations} 
where $d{\bf a}$ is the vector surface element given by $d{\bf a}={\bf n}da$. Substituting Eqs. (\ref{eq-expansion_EM1}) and (\ref{eq-expansion_EM2}) into Maxwell's equations given by 
\begin{subequations} 
\begin{align} 
&{\bm\nabla}\times{\bf E}+\frac{1}{c}\frac{\partial{\bf B}}{\partial t}={\bf 0}, 
\label{eq-Faraday-2} 
\\
&{\bm\nabla}\times{\bf B}-\frac{1}{c}\frac{\partial{\bf E}}{\partial t}=\frac{4\pi}{c}{\bf j}, 
\label{eq-Ampere}, 
\\
&{\bm\nabla}\cdot{\bf B}=0 
\label{eq-no_magnetic_monopole-2}, 
\\
&{\bm\nabla}\cdot{\bf E}=4\pi\rho 
\label{eq-Gauss-2},
\end{align}
\end{subequations} 
we then obtain 
\begin{subequations} 
\begin{align} 
k_i\int{\bf E}\cdot{\bf e}_i
dv+\frac{1}{c}\frac{\partial}{\partial t}\int{\bf B}\cdot{\bf b}_i
dv 
&=-\int_S({\bf E}\times{\bf b}_i)\cdot
d{\bf a}, \label{eq-Faraday} 
\\
k_i\int{\bf B}\cdot{\bf b}_i
dv-\frac{1}{c}\frac{\partial}{\partial t}\int{\bf E}\cdot{\bf e}_i
dv 
&=\frac{4\pi}{c}\int{\bf j}\cdot{\bf e}_i\,dv 
-\int_{S'}({\bf B}\times{\bf e}_i)\cdot
d{\bf a}, \label{eq-Ampere2} 
\\
-\frac{1}{c}\frac{\partial}{\partial t}\int{\bf E}\cdot{\bf f}_i
dv 
&=\frac{4\pi}{c}\int{\bf j}\cdot{\bf f}_i
dv, \label{eq-no_magnetic_monopole} 
\\
-k_i\int{\bf E}\cdot{\bf f}_i
dv \label{eq-Gauss}
&=4\pi\int\rho\psi_i dv, 
\end{align} 
\end{subequations} 
where $c$ is the speed of light in vacuum. Using Eqs. (\ref{eq-no_magnetic_monopole}) and (\ref{eq-Gauss}), we can show the equation of continuity given by 
\begin{align} 
\frac{\partial\rho}{\partial t}+{\bm\nabla}\cdot{\bf j}=0. 
\end{align}
We also obtain separate equations for $\int{\bf E}\cdot{\bf e}_i dv$ and $\int{\bf B}\cdot{\bf b}_i dv$ from Eqs. (\ref{eq-Faraday}) and (\ref{eq-Ampere2}) as 
\begin{subequations} 
\begin{align} 
&\frac{1}{c^2}\frac{\partial^2}{\partial t^2}\int{\bf E}\cdot{\bf e}_i
dv 
+k_i^2\int{\bf E}\cdot{\bf e}_i
dv
=-\frac{1}{c}\frac{\partial}{\partial t}
\left[\frac{4\pi}{c}\int{\bf j}\cdot{\bf e}_i
dv-\int_{S'}({\bf B}\times{\bf e}_i)\cdot
d{\bf a}\right]
-k_i\int_S({\bf E}\times{\bf b}_i)\cdot
d{\bf a}, \label{eq-cavityE} 
\\
&\frac{1}{c^2}\frac{\partial^2}{\partial t^2}\int{\bf B}\cdot{\bf b}_i dv 
+k_i^2\int{\bf B}\cdot{\bf b}_i dv
=k_i\left[\frac{4\pi}{c}\int{\bf j}\cdot{\bf e}_i
dv-\int_{S'}({\bf B}\times{\bf e}_i)\cdot
d{\bf a}\right] 
-\frac{1}{c}\frac{\partial}{\partial t}\int_S({\bf E}\times{\bf b}_i)\cdot
d{\bf a}. 
\end{align}\label{eq-cavity}%
\end{subequations} 
We can obtain the expansion coefficients of the EM fields in a hollow cavity by solving these wave equations. 

\section{Solution for the nonlinear London equation and surface reactance\label{app:c}} 

With the same procedures in App. B of Ref. \cite{xu95}, we obtain a solution for the nonlinear London equation, and then, with the same procedures in Ref. \cite{zar23}, we also reproduce the surface reactance due to the NLM effect of an SRF cavity in Bogorad's setup \cite{bog19}. 

In the absence of phase vortices, we absorb the phase gradient ${\bm\nabla}\vartheta$ into the vector potential, ${\bf A}+(\hbar c/2e){\bm\nabla}\vartheta\to{\bf A}$, and work in the transverse gauge, ${\bm\nabla}\cdot{\bf A}=0$. Then, the nonlinear London equation is given by 
\begin{align} 
{\bm\nabla}^2{\bf A}
-\frac{1}{\lambda_{\rm L}^2}
\left[ 
1
-\theta
\left(
\frac{A}{A_c}
\right)^2
\right]{\bf A}
=
{\bf 0}, 
\end{align}
with $A_c\equiv H_c\lambda_{\rm L}$. We consider the geometry in which the superconductor occupies the half-space $x>0$. The surface field ${\bf H}=H\hat{\bf y}$ is parallel to the interface, and vector potential ${\bf A}=A\hat{\bf z}$ is parallel to $z$. The field in the superconductor is given by ${\bf B}=(dA/dx)\hat{\bf y}$. Continuity of the parallel filed at the interface imposes the following boundary condition:
\begin{align} 
\frac{dA}{dx}\bigg|_{x=0}=H. 
\end{align}
The field and vector potential vanish deep inside the superconductor. We introduce a dimensionless vector potential $a\equiv\sqrt{\theta}(A/A_c)$ and distance $\chi\equiv x/\lambda_{\rm L}$, and then the differential equation for $a(\chi)$ is given by
\begin{align} 
\frac{d^2a}{d\chi^2}-(1-a^2)a=0, \label{eq-a}
\end{align}
with the following boundary conditions: 
\begin{align} 
\frac{da}{d\chi}\bigg|_{\chi=0}=\sqrt{\theta}\frac{H}{H_c}\equiv h, 
\ \ \ a(\chi\to\infty)=0. \label{eq-BC-a} 
\end{align}
By multiplying Eq. (\ref{eq-a}) by $da/d\chi$ and integrating the resulting equation with the asymptotic boundary $a(\chi\to\infty)=0$, we obtain a first integral of Eq. (\ref{eq-a}) as 
\begin{align} 
\frac{da}{d\chi}=-a\sqrt{1-\frac{a^2}{2}}. 
\end{align}
In the physically relevant limit $|a|\ll1$, we replace $\sqrt{1-a^2/2}$ to $1-a^2/4$, integrate the resulting equation, and then obtain the dimensionless vector potential $a$ as 
\begin{align} 
a=\frac{a_0}{\sqrt{a_0^2/4+(1-a_0^2/4){\rm e}^{2\chi}}}, 
\end{align}
with $a_0\equiv a(\chi=0)$. 
Using it, we also obtain the dimensionless magnetic field as 
\begin{align} 
b\equiv\frac{da}{d\chi}
=-\frac{a_0(1-a_0^2/4){\rm e}^{2\chi}}{[a_0^2/4+(1-a_0^2/4){\rm e}^{2\chi}]^{3/2}}.
\label{eq-b}
\end{align}
The constant $a_0$ is determined by the boundary condition [Eq. (\ref{eq-BC-a})] as $h=-a_0(1-a_0^2/4)$, and then Eq. (\ref{eq-b}) is rewritten as 
\begin{align} 
b =\frac{h{\rm e}^{2\chi}}{[h^2/4+(1-h^2/4){\rm e}^{2\chi}]^{3/2}}.
\label{eq-b2}
\end{align}
For weak nonlinearity, $a_0$ is also given by 
\begin{align} 
a_0=\int_\infty^0d\chi b
\approx
-h
\left(
1+\frac{h^2}{4}
\right), 
\end{align}
and then $a$ is expressed with $h$ as 
\begin{align} 
a=-\frac{h(1+h^2/4)}{\sqrt{h^2/4+(1-h^2/4){\rm e}^{2\chi}}}. 
\end{align}
Therefore, we obtain the vector potential up to the first-order in $\theta$ as 
\begin{align} 
A\approx-H\lambda_{\rm L}{\rm e}^{-x/\lambda_{\rm L}}
\left[
1+\frac{\theta}{8}\left(\frac{H}{H_c}\right)^2(3-{\rm e}^{-2x/\lambda_{\rm L}})
\right]. \label{eq-A}
\end{align}

We consider the same system in Sec.~\ref{sec:ii} and use Eq. (\ref{eq-H-NLM}) for the surface magnetic field. Expressing the vector potential as 
\begin{align} 
A_{{\rm NLM}z}(x)=&{\rm Re}A_{{\rm NLM}z}(x){\rm e}^{-\frac{\omega_3^\infty}{2Q_3}t}\cos\omega_{\rm NLM}t
+{\rm Im}A_{{\rm NLM}z}(x){\rm e}^{-\frac{\omega_3^\infty}{2Q_3}t}\sin\omega_{\rm NLM}t, 
\end{align}
and using Eq. (\ref{eq-A}) replacing $x$ with $x+R$, we obtain $A_{{\rm NLM}z}$ as
\begin{align} 
&A_{{\rm NLM}z}(x)=
-B_{{\rm NLM}\varphi}(R)\lambda_{\rm L}{\rm e}^{-x'/\lambda_{\rm L}}
\times
\left[
1+\frac{\theta}{32H_c^2}
\frac{1-{\rm e}^{-\frac{3\omega_3^\infty}{2Q_3}t_p}}{1-{\rm e}^{-\frac{\omega_3^\infty}{2Q_3}t_p}}
|B_{{\rm NLM}\varphi}(R)|^2
\left(
3-{\rm e}^{-2x'/\lambda_{\rm L}}
\right)
\right], 
\end{align}
with $x'=x-R$. We can obtain an equation for the frequency shift at low temperatures from the boundary conditions and its derivative at the vacuum-superconductor interface. The NLM signal vector potential on the vacuum side is the TM$_{020}$ mode given by $A_{{\rm NLM}z}(x)=A_0J_0(\omega_{\rm NLM}x/c)$. Then, the ratio of the two boundary conditions reduces to 
\begin{align} 
&-\frac{dA_{{\rm NLM}z}(x)/dx|_{x=R}}{A_{{\rm NLM}z}(x=R)}
=\frac{\omega_{\rm NLM}}{c}\frac{J_1(\omega_{\rm NLM}R/c)}{J_0(\omega_{\rm NLM}R/c)}
=\frac{1}{\lambda_{\rm L}}
\left[
1+\frac{\theta}{16H_c^2}
\frac{1-{\rm e}^{-\frac{3\omega_3^\infty}{2Q_3}t_p}}{1-{\rm e}^{-\frac{\omega_3^\infty}{2Q_3}t_p}}
|B_{{\rm NLM}\varphi}(R)|^2
\right]^{-1}. \label{eq-omega_NLM}
\end{align}
We can obtain the frequency shift relative to $\omega_3^\infty$ due to the NLM effect by solving Eq. (\ref{eq-omega_NLM}) with Eq. (\ref{eq-surface_magnetic_field-NLM}) directly.

We calculate the surface reactance using Eq. (\ref{eq-omega_NLM}). The corresponding EM fields are given by $E_{{\rm NLM}z}(x)=(i\omega_{\rm NLM}/c)A_0J_0(\omega_{\rm NLM}x/c)$ and $B_{{\rm NLM}\varphi}(x)=(\omega_{\rm NLM}/c)A_0J_1(\omega_{\rm NLM}x/c)$ (See also App. \ref{app:a}). Using these NLM EM fields and Eqs. (\ref{eq-Zs-2}) and (\ref{eq-omega_NLM}), we obtain the surface reactance at low temperatures as 
\begin{align} 
&X_{\rm NLM}^s 
=Z_0\frac{\omega_{\rm NLM}}{c}\lambda_{\rm L}
\times
\left[
1+
\frac{\theta}{16H_c^2}
\frac{1-{\rm e}^{-\frac{3\omega_3^\infty}{2Q_3}t_p}}{1-{\rm e}^{-\frac{\omega_3^\infty}{2Q_3}t_p}}
|B_{{\rm NLM}\varphi}(R)|^2
\right]. 
\end{align}
Noting that the frequency shift in the Caucy-Lorentz distribution cannot be neglected even if $ \delta\omega_{\rm NLM}\ll\omega_3^\infty$, we can replace $\omega_{\rm NLM}$ in the prefactor with $\omega_3^\infty$ and then reproduce Eq. (\ref{eq-Xs}). 

%
\end{document}